\documentclass[twocolumn]{openjournal}

\usepackage{xcolor}
\usepackage{textgreek}
\usepackage{amsmath}
\usepackage[utf8]{inputenc}
\usepackage[english]{babel}

\usepackage{hyperref}
\hypersetup{
    unicode, 
    colorlinks=true,
    linkcolor=linkcolor,
    citecolor=linkcolor,
    filecolor=linkcolor,
    urlcolor=linkcolor,
}
\usepackage{color,colortbl}
\definecolor{linkcolor}{rgb}{0.0,0.3,0.5}
\usepackage{tensind}
\tensordelimiter{?}
\DeclareGraphicsExtensions{.bmp,.png,.jpg,.pdf}
\usepackage{verbatim}
\usepackage[normalem]{ulem}
\usepackage{orcidlink}
\usepackage{soul}

\begin{document}
\title{Search for neutrino emission from blazar $\gamma$-ray flares accounting for possible neutrino time delays}

\author{Egor Podlesnyi\orcidlink{0000-0003-3395-0419}}
\email{egor.podlesnyi@ntnu.no}
\author{Foteini Oikonomou\orcidlink{0000-0002-0525-3758}}
\affiliation{Department of Physics, Norwegian University of Science and Technology, Høgskoleringen 5, Trondheim 7491, Norway}

\begin{abstract}
    We report the results of the search for the high-energy neutrino emission associated with blazar flares, accounting for a possible lag of neutrinos with respect to the electromagnetic emission, either due to the slowness of the proton energy losses in $p\gamma$ collisions and/or proton acceleration. We perform two tests, cross-matching neutrinos with energies $E_{\nu} \gtrsim 100$~TeV from the public catalogue of neutrino alerts IceCat-1 with active galactic nuclei from two source samples based on 1)~the MOJAVE database and 2)~the CGRaBS catalogue, and utilising \textit{Fermi}-LAT light curves from the public light curve repository. We scan over a wide range of values of the jet-frame time delay $t^{\prime}_{\mathrm{delay}}$ between the neutrino arrival and the time of the prior major $\gamma$-ray flare and find a pre-trial $\sim 2\sigma$ correlation at $t^{\prime}_{\mathrm{delay}} \sim 10^{3}$~d, which is consistent ($p_{\mathrm{post-trial}} \sim 0.1$) with expectations under the null hypothesis after trial correction.
\end{abstract}


\maketitle

\section{Introduction}\label{sec:intoduction}
    In the 2010s, the \textit{IceCube} neutrino observatory discovered astrophysical high-energy ($E_{\nu} \gtrsim 1$~TeV) neutrinos \citep{IceCube:2013low}. Around $\sim 5-30$\% of their flux was later attributed to the Milky Way galaxy \citep{IceCube:2023ame,Kovalev:2022izi,Baikal-GVD:2024kfx}. Aside from that, besides the hint of an individual association of the multi-wavelength flare in blazar TXS~0506+056 with a $\gtrsim 200$~TeV neutrino \citep{IceCube:2018dnn} and hints of $0.3-100$~TeV neutrino emission from active galaxies NGC 1068 \citep{IceCube:2022der}, NGC 3079 and NGC 4151 \citep{Neronov:2023aks}, the main contributors to the observed neutrino flux remain unknown (for a recent review, see, e.g., \citealp{Troitsky:2021nvu,Troitsky:2023nli,Kurahashi:2022utm}).

    Active galactic nuclei (AGNs) and their subset with relativistic jets pointing to the Earth at a small angle, blazars, were proposed as neutrino sources almost half a century ago \citep{Berezinsky81a}. Still, it was not until recently when some hints of their detectable contribution to the observed high-energy neutrino flux have been found \citep{Plavin:2020emb,Plavin:2020mkf,Plavin:2022oyy,Plavin:2023wsb,Plavin:2025pjt,Hovatta:2020lor,Giommi:2020hbx,Kouch:2024xtd,Buson:2022fyf,Buson:2022fyfErratum,Suray:2023lsa,ANTARES:2023lck}. Yet, the (in)significance of the blazar-neutrino association depends on neutrino datasets, blazar catalogues and test statistics used in searches for correlations between AGNs and neutrinos \citep{Bellenghi:2023yza,IceCube:2023htm}.
    
    Previous studies, looking for associations between blazars and \textit{IceCube} high-energy neutrinos, used as proxies either time-averaged fluxes of sources in various electromagnetic bands (from radio to $\gamma$-ray) or their light curves with corresponding values close to the time of the neutrino arrival. However, from the theoretical point of view, the temporal coincidence of the neutrino arrival and the major electromagnetic flare\footnote{Here, we focus on $\gamma$-ray flares; note that radio flares tend to lag behind $\gamma$-ray flares \citep{Kramarenko:2021rkf} and have longer duration, so we do not discuss them here; optical flares tend, in principle, happen together with $\gamma$-ray flares (although, there might be so-called orphan flares, see, e.g. \citealp{Wang:2021unc}), but there are no continuous observations of a large sample of AGNs with no gaps in observations, which makes it difficult to achieve a potentially significant correlation as demonstrated by \citet{Kouch:2025ipk}.} might not be warranted for at least two reasons: (i) slow proton acceleration, and/or (ii) slow proton energy losses.
    
    The time required to accelerate protons to energies corresponding to the energies ($\gtrsim 10^{14}$~eV) of neutrinos produced in $p\gamma$ collisions is much larger than the acceleration time of electrons responsible for electromagnetic flares. Given that a neutrino with energy $E_{\nu}$ is typically produced by a proton with energy $E_{p} \sim 20 E_{\nu}$, and that the energy of the neutrino produced in a jetted source with Doppler factor $D$ and redshift $z$ in the jet frame\footnote{Throughout the text, the primed $^{\prime}$ quantities are defined in the rest frame of the jet plasma bulk motion of the corresponding source (shortly, the jet frame).} is $E_{\nu}^{\prime} = E_{\nu} (1 + z) / D$, we obtain $E_{\nu} \sim E_{p}^{\prime} D / (20[1 + z])$, i.e. the observed neutrino energy $E_{\nu}$ is roughly equal to the jet-frame parent proton energy $E_{p}^{\prime}$ for typical values of $D \sim 30$, $z \sim 0.5$. For most blazars, their electromagnetic emission during flares can be successfully explained with leptonic models with electrons reaching energies $E^{\prime \max}_{e} \sim 10^{9}-10^{13}$~eV (depending on the blazar class and model assumptions, see, e.g., \citealp{Ghisellini:2009wa, Ghisellini:2009fj, Zech:2021emy, Rodrigues:2025cpm}). If the protons are accelerated in the same zone as the electrons via the same mechanism, then, e.g., in the assumption of the linear dependence of the acceleration timescale on the maximum energy, $t_{\mathrm{acc}}^{\prime} \propto E^{\prime \max}$, protons will take one to five orders of magnitude longer time to accelerate to energies required for the neutrino production than electrons to reach energies necessary for explaining the flaring $\gamma$-ray emission.
    
    This difference is explained by the fact that the maximum electron energy is defined by the electron acceleration time reaching the characteristic energy-loss time (either due to synchrotron or inverse Compton losses), which is typically much shorter than the characteristic escape time from the emitting region. On the contrary for protons, given that at energies $\sim 10^{14}$~eV AGN jets are optically thin for them in most scenarios (both for proton-synchrotron\footnote{Indeed, at a fixed energy, synchrotron energy-loss timescale $t_{\mathrm{syn}} = E / (-dE/dt)$ \citep{Blumenthal:1970gc} for protons is larger than for electrons by $(m_p / m_e)^{4} \sim 10^{13}$ so the increase of the proton energy even by a factor of $10^{5}$ cannot compensate for this dramatic difference given the linear dependence of $t_{\mathrm{syn}}$ on the particle energy.} and $p\gamma$ radiative processes), they are capable of reaching much higher energies due to low energy losses and high energy thresholds of $p\gamma$ interactions. Thus, protons have their maximum energy defined either by their escape time, i.e. by the Hillas criterion \citep{Hillas:1984ijl}, which is not applicable for electrons due to the optical thickness of the emitting region for them, or by the $p\gamma$ energy-loss timescale, which is much larger than the characteristic timescale of the electron energy losses. This difference is clearly illustrated in Fig.~1 of \citet{Rodrigues:2025cpm} and Fig.~12 of \citet{Podlesnyi:2025aqb}, where an orders-of-magnitude difference in both maxima energies and acceleration timescales (up to these maxima energies) of electrons and protons is shown.
    
    Thus, during major flares in blazars, injected electrons typically cool within days or less in the observer frame, but protons need a much longer time to interact with photons for effective production of neutrinos if it happens in the energy range of $\sim 100$~TeV near the threshold of the photopion production on optical/ultraviolet/X-ray photon fields.
    
    Whether it is the acceleration timescale $t^{\prime}_{\mathrm{acc}}$ or the $p\gamma$ energy-loss timescale $t^{\prime}_{p\gamma}$ which causes the possible time lag $t^{\prime}_{\mathrm{delay}}$ of the neutrino arrival depends on the conditions in the source, such as the acceleration mechanism, magnetic field, and target photon fields, and it is the slowest process which will determine the neutrino time lag, i.e. $t^{\prime}_{\mathrm{delay}} \sim \max \{t^{\prime}_{\mathrm{acc}}, \, t^{\prime}_{p\gamma}\}$. Depending on the model parameters, $\gtrsim 100$~TeV neutrinos can arrive months to years later than the corresponding electromagnetic flare peak \citep{Podlesnyi:2025aqb}. A study of a possible association of a few lagging neutrinos with another class of sources, namely tidal disruption events (TDEs), was recently performed by \citet{Lu:2025vmk} (see, as well, the recent studies by \citet{IceCube:2025uzh} and \citet{Langis:2026dyj}, where neutrino-TDEs correlations are found to be not significant).

    If one argues that the proton acceleration timescale is not substantially larger than the electron acceleration timescale, and the $p\gamma$ collisions happen far from the threshold and efficiently, then protons can be accelerated to energies as high as $\sim 100$~PeV and produce neutrinos of corresponding energies (with a negligible time delay with respect to the electromagnetic flare), since the peak energy of the neutrino spectral-energy distribution closely follows the maximum proton energy \citep{Podlesnyi:2025aqb, Rodrigues:2025cpm}. However, only one neutrino with energy $E_{\nu}\gtrsim$100 PeV has been discovered so far, namely KM3-230213A~\citep{KM3NeT2025}. Thus, some of those $\gtrsim 100$~TeV astrophysical neutrinos detected by the \textit{IceCube} observatory \citep{IceCube:2023agq}, if they are produced in a causal connection with major flares in blazars, can be expected to come with some time delay with respect to the flare peaks, either due to the large proton acceleration or $p\gamma$ energy-loss timescale.

    The publicly available \textit{Fermi}-LAT light-curve repository (LCR) \citep{Fermi-LAT:2023iml} with regular-binned light curves of all significantly variable \textit{Fermi}-LAT-detected AGNs spanning more than 15 years allows us to search for association of major $\gamma$-ray flares with lagging behind them \textit{IceCube} high-energy neutrinos published by the \textit{IceCube} collaboration in the first alert catalogue IceCat-1 \citep{IceCube:2023agq,IceCat-1v4}. In this paper, we make the first attempt at this search. In Sect.~\ref{sec:data}, we present the datasets and catalogues used in our work; in Sect.~\ref{sec:methods}, we present the methods used for searching for associations of lagging neutrinos with major $\gamma$-ray flares in AGNs; in Sect.~\ref{sec:results}, the results are shown, and caveats of our analysis are highlighted in Sect.~\ref{sec:caveats}. We discuss implications of our findings and conclude in Sect.~\ref{sec:discussion}.

\section{Data}\label{sec:data}
    \subsection{Neutrino alerts from the IceCat-1 catalogue}\label{sec:icecat}
    We use the IceCat-1 catalogue of high-energy neutrino alerts published by the \textit{IceCube} collaboration \citep{IceCube:2023agq,IceCat-1v4}. In the following, we use the terms ``alert'' and ``neutrino'' interchangeably. We remove eight entries for which there is a flag indicating a coinciding signal in the \textit{IceTop} detector of cosmic rays, implying a high chance of those alerts being background. This leaves us with $N_{\nu} = 348$ neutrino alerts. For each alert $\nu$, the following information of interest is available:
    \begin{equation}
        x_{\nu} = \{s_{\nu}, t_{\nu}, E_{\nu}, \alpha_{\nu}, \delta_{\nu}, \vec{\epsilon}_{\nu} \},
    \end{equation}
    where $s_{\nu}$ is the neutrino signalness reported by the \textit{IceCube} collaboration in the IceCat-1 catalogue which varies from $0.126$ to $0.997$ with a median value of $0.410$ and represents the probability of the neutrino $\nu$ being of astrophysical origin assuming an $E^{-2.19}_{\nu}$ astrophysical neutrino power-law flux (although this assumption does not hold, e.g., in cases when neutrinos are produced in $p\gamma$ interactions near the threshold, see, e.g., \citealp{Rodrigues:2024fhu, Rodrigues:2023vbv, Kuhlmann:2025ocn}), $t_{\nu}$ is the moment of the neutrino detection, $E_{\nu}$ is the estimated muon energy, which is a good proxy of the neutrino energy (see Fig.~7 by \citealp{IceCube:2023agq}), $ \alpha_{\nu}$ is the right ascension (RA), and $\delta_{\nu}$ is the declination of the most probable sky coordinate of the neutrino arrival, while the vector $\vec{\epsilon}_{\nu} = \{\alpha_{\nu}^{+}, \, \alpha_{\nu}^{-}, \, \delta_{\nu}^{+}, \, \delta_{\nu}^{-} \}$ incorporates information about the uncertainty of the reconstructed direction of the neutrino arrival, where $(\alpha_{\nu}^{+} + \alpha_{\nu}^{-})$ defines the width, and $(\delta_{\nu}^{+} + \delta_{\nu}^{-})$ defines the length of the minimum rectangle encapsulating the neutrino uncertainty arrival contours (see Fig.~1 by \citealp{IceCube:2023htm}).

    \subsubsection{Set of mock neutrino datasets}\label{sec:mock}
        Based on the alert dataset $\{ x_{\nu} \}$, we create a set of $N_{f} = 49 \, 999$ mock datasets $\{ m_{\nu}^{f} \}$, $f = 1, \,..., \, N_{f}$, where for each mock alert all the parameters $x_{\nu}$ remained unchanged except for the RA $\alpha_{\nu}^{f}$ and the arrival time $t_{\nu}^{f}$, i.e.
        \begin{equation}
            m_{\nu}^{f} = \{s_{\nu}, t_{\nu}^{f}, E_{\nu}, \alpha_{\nu}^{f}, \delta_{\nu}, \vec{\epsilon}_{\nu}\}.
        \end{equation}
        Since the \textit{IceCube} detector is located at the South Pole, its sensitivity to astrophysical neutrinos on a long timescale depends only on the energy and declination. Thus, following \citet{Plavin:2020emb}, by generating a random uniformly distributed number between $0^{\circ}$ and $360^{\circ}$ for each of the values of $\alpha_{\nu}^{f}$, we obtain a set of mock neutrino datasets reflecting a situation of neutrinos registered from random sky directions (losing a possible connection with a potential class of sources) but keeping the dependence of the \textit{IceCube} sensitivity on the declination unchanged. For correlation searches involving the time of the neutrino arrival, we generate random values of $t_{\nu}^{f}$ from a uniform distribution $U[t_{\nu}^{\mathrm{min}},t_{\nu}^{\mathrm{max}}]$, where $t_{\nu}^{\mathrm{min}} = 55695.064$ (MJD) is the arrival time of the first neutrino in the IceCat-1, and $t_{\nu}^{\mathrm{max}} = 60231.917$ (MJD) is the time of the last alert in IceCat-1. Thus, the total time span of IceCat-1 is $T_{\mathrm{\textit{IceCube}}} \approx 4537$~d~$\approx 12.4$~yr. Using the Kolmogorov-Smirnov test, we verified that the original set of RAs of \textit{IceCube} neutrinos $\{ \alpha_{\nu} \}$ and their arrival times $\{ t_{\nu} \}$ cannot be distinguished from a set drawn from the corresponding uniform distribution.

    \subsection{Source lists}\label{sec:blazars}
    \subsubsection{MOJAVE-based source list}\label{sec:MOJAVE}
    Since the aim of the paper is to search for blazar-neutrino associations with account for a possible neutrino time lag, which we assume to be universal in the jet frame, to transform the anticipated neutrino arrival delay from the jet frame into the observer frame, the measurements of the blazar's Doppler factor $D$ and redshift $z$ are required, since $t_{\mathrm{\, delay}} = t^{\prime}_{\mathrm{delay}} (1 + z) / D$. As a starting point, we use the sample of $447$ AGNs compiled by \citet{Homan:2021ijm}, which consists of radio-bright AGNs monitored by the MOJAVE program. All sources in the selected list are located at declinations $> -30^{\circ}$ and have a minimum 15~GHz correlated flux density $\gtrsim 50$~mJy. All these sources are also present in the Radio Fundamental Catalogue (RFC), version 2025a \citep{Petrov2025ApJS}. We utilize Table 4 in the work of \citet{Homan:2021ijm}, where the Doppler factors of the blazars are reported, obtained under the assumption that all sources have observed brightness temperature $\propto D T^{\prime}_{b} / (1 + z)$, and the intrinsic brightness temperatures $T^{\prime}_{b}$ are determined according to \citet{Kovalev:2005yd}.
    
    As the next step, we use the 4FGL-DR4 catalogue of \textit{Fermi}-LAT sources \citep{Fermi-LAT:2019yla} available at the mission website\footnote{\url{https://fermi.gsfc.nasa.gov/ssc/data/access/lat/14yr_catalog/gll_psc_v35.fit}} as the file \texttt{gll\_psc\_v35.fit}. We cross-match 4FGL sources with those presented in the catalogue of \citet{Homan:2021ijm}.

    We use the 4LAC catalogue \citep{Fermi-LAT:2019pir} to obtain the redshift values for the sources in our list of blazars. For those objects which do not have a finite redshift value in the 4LAC catalogue, we use the \texttt{astroquery.ned} module to query the redshift from the NASA/IPAC Extragalactic Database \citep{NED}. For blazars for which only lower limits on Doppler factors were reported by \citet{Homan:2021ijm} due to unknown redshift, we multiply the lower limits on $D$ by $(1 + z)$ with the values of $z$ we found in the 4LAC catalogue or the NASA/IPAC Extragalactic Database\footnote{This is because the observed brightness temperature $\propto D T^{\prime}_{b} / (1 + z)$ was assumed by \citet{Homan:2021ijm} to correspond to $z = 0$ for sources with unknown redshifts, hence the lower limit on $D$ was reported. Among the sources of \citet{Homan:2021ijm} associated with a \textit{Fermi}-LAT source, (i) there are 43 sources with Doppler factors with lower limits due to unknown redshifts for which we have found the redshifts; (ii) there are 22 sources with Doppler factors with lower limits \textit{not} due to unknown redshifts --- these sources are removed from the further analysis.}.
    
    Finally, we use the \textit{Fermi}-LAT LCR \citep{Fermi-LAT:2023iml} and download monthly-binned photon flux light curves with a minimum detection significance of $\mathrm{TS} = 4$ $(2 \sigma)$ required in each time bin and a spectral fitting with a fixed photon index (default light curves). We leave only those objects which are present in the LCR, i.e. those with significant variability. The final list of 4FGL AGNs with known redshifts, $\gamma$-ray light curves, and Doppler factors reported by \citet{Homan:2021ijm} consists of $N_b = 294$ sources. Among them, according to the classification in the 4FGL catalogue, there are 184 flat-spectrum radio quasars (FSRQs), 93 BL Lacertae-like objects (BLLs), 6 radio galaxies (RDGs), 4 blazars of uncertain type (BCUs), 4 Seyfert galaxies of type 1 (NLSY1s), 2 compact steep-spectrum radio sources (CSSs), and 1 steep-spectrum radio quasar (SSRQ). Since most of the sources are blazars, we use the terms ``source'', ``AGN'', and ``blazar'' interchangeably throughout the paper. The obtained blazar list is denoted as H21+.

    \subsubsection{CGRaBS-based source list}

    \citet{Homan:2021ijm} obtained Doppler factors under the assumption that the apparent brightness temperature at $15$~GHz depends only on the blazar jet Doppler factor $D$ and the intrinsic (jet-frame) brightness temperature, which for all blazars in the sample is equal to $4 \times 10^{10}$~K, as found according to \citet{Kovalev:2005yd}. This approach can give only approximate values of the Doppler factor because it does not allow for intrinsic source-to-source variations and, besides, may result in incorrect estimations of $D$ when radio images of the jet and the core of the blazar are superimposed \citep{Kovalev:2025kxf}.

    Thus, it is of interest to search for blazar-neutrino associations using a sample of sources with their Doppler factors obtained with a different method. Recently, \citet{Rodrigues:2023vbv} modelled time-averaged spectral-energy distributions (SEDs) for a sample of 324 blazars, obtaining a fit to the multi-wavelength SEDs with a leptonic and a leptohadronic model, determining values of the Doppler factor $D$ (among other parameters) for each source. The source sample of \citet{Rodrigues:2023vbv} is based on the flux-limited ($F_{8.4 \mathrm{\, GHz}} > 65$~mJy) CGRaBS catalogue of radio-loud AGNs \citep{Healey:2007gb} which was cross-matched by \citet{Paliya:2017xaq} with sources from various \textit{Fermi}-LAT catalogues.
    
    We cross-match sources from the list of 324 blazars \citep{Rodrigues:2023vbv}\footnote{Their Doppler factors and redshifts available at \url{https://github.com/xrod/lephad-blazars/blob/main/model_parameters.csv}.} with the \textit{Fermi}-LAT LCR \citep{Fermi-LAT:2023iml}. This results in $N_b = 295$ sources with known $D$, $z$ and light curves in the LCR. Among them, according to the \textit{Fermi}-LAT classification \citep{Fermi-LAT:2019yla}, there are $218$ FSRQs, $72$ BLLs, $2$ BCUs, and $3$ NLSY1s. We denote the compiled source list as R24+.

    \subsection{Properties of the source lists}\label{sec:properties}
    \begin{figure}
        \centering
        \includegraphics[width=0.49\textwidth]{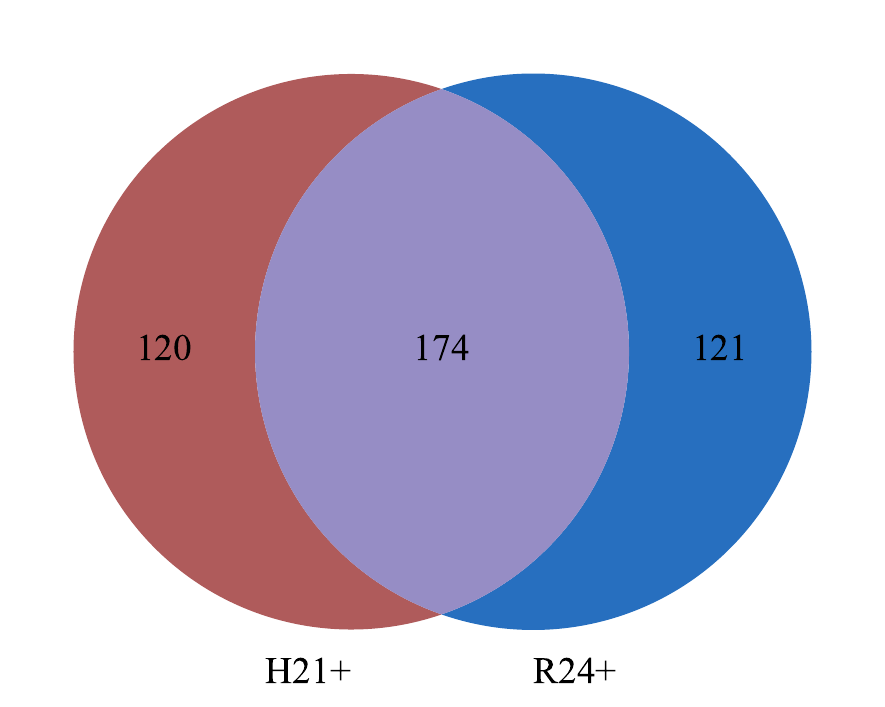}
        \caption{Venn diagram showing the number of sources in H21+ and R24+ source lists and the number of overlapping sources.\label{fig:Venn}}
    \end{figure}
    \begin{figure}
        \centering
        \includegraphics[width=0.49\textwidth]{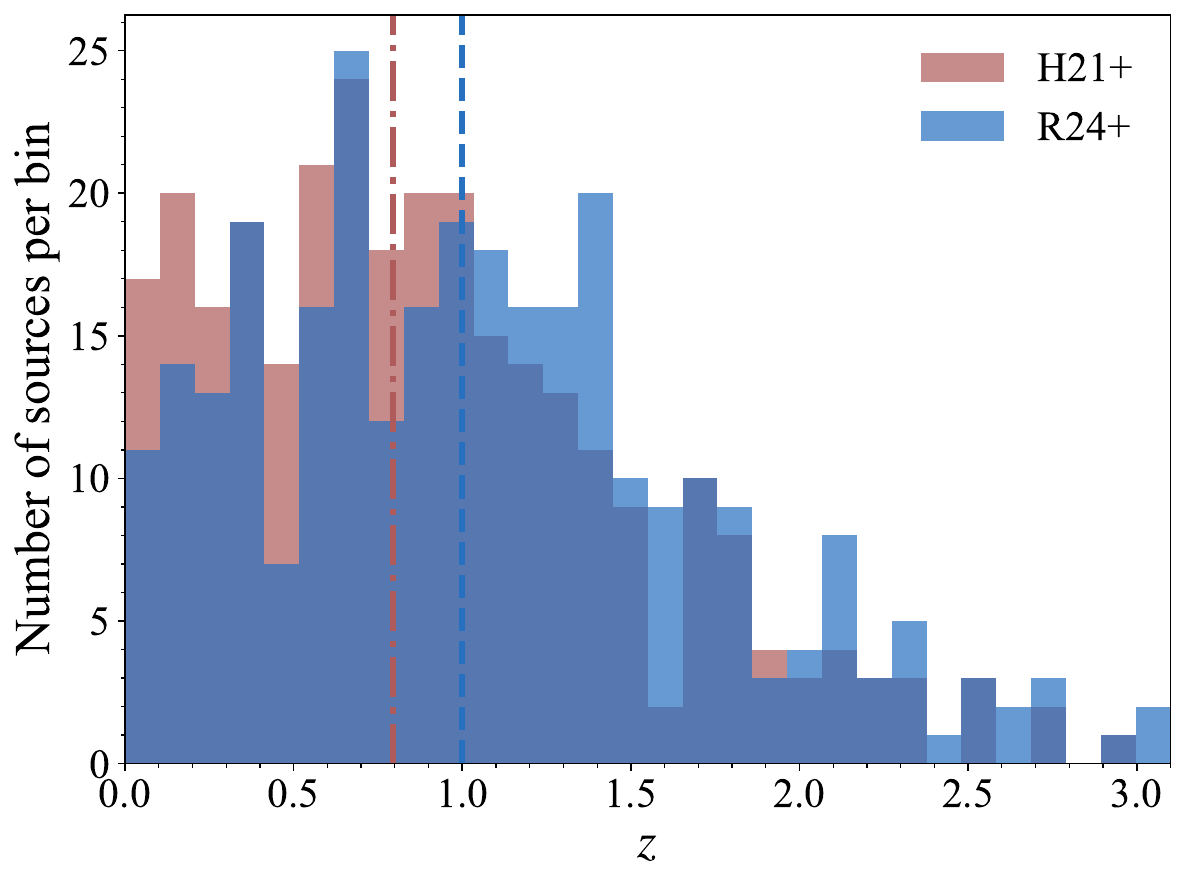}
        \caption{Redshift histograms for H21+ and R24+ source lists. The median redshifts are shown as dashed-dotted and dashed vertical lines for the former and the latter, respectively.\label{fig:redshift}}
    \end{figure}
    \begin{figure}
        \centering
        \includegraphics[width=0.49\textwidth]{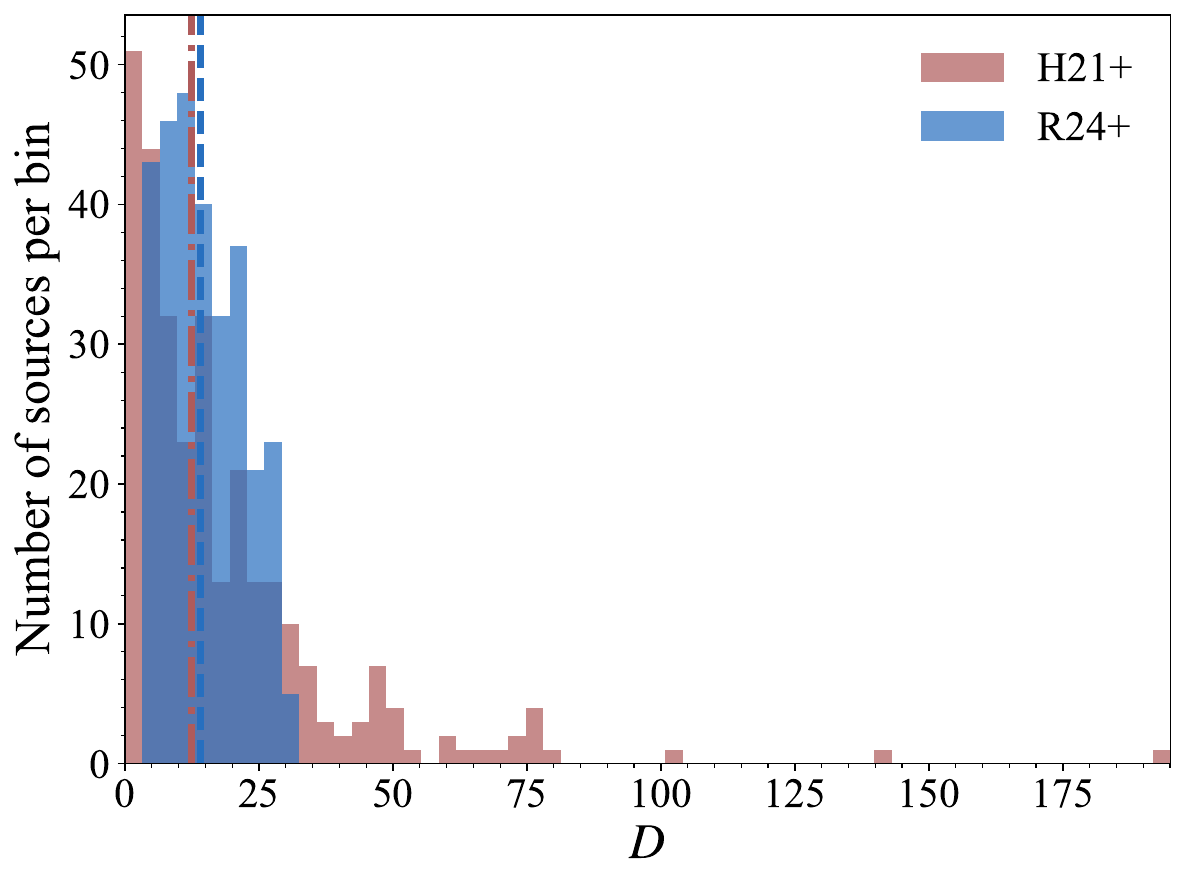}
        \caption{Same as Fig.~\ref{fig:redshift} but for Doppler factors $D$.\label{fig:doppler}}
    \end{figure}
    \begin{figure}
        \centering
        \includegraphics[width=0.49\textwidth]{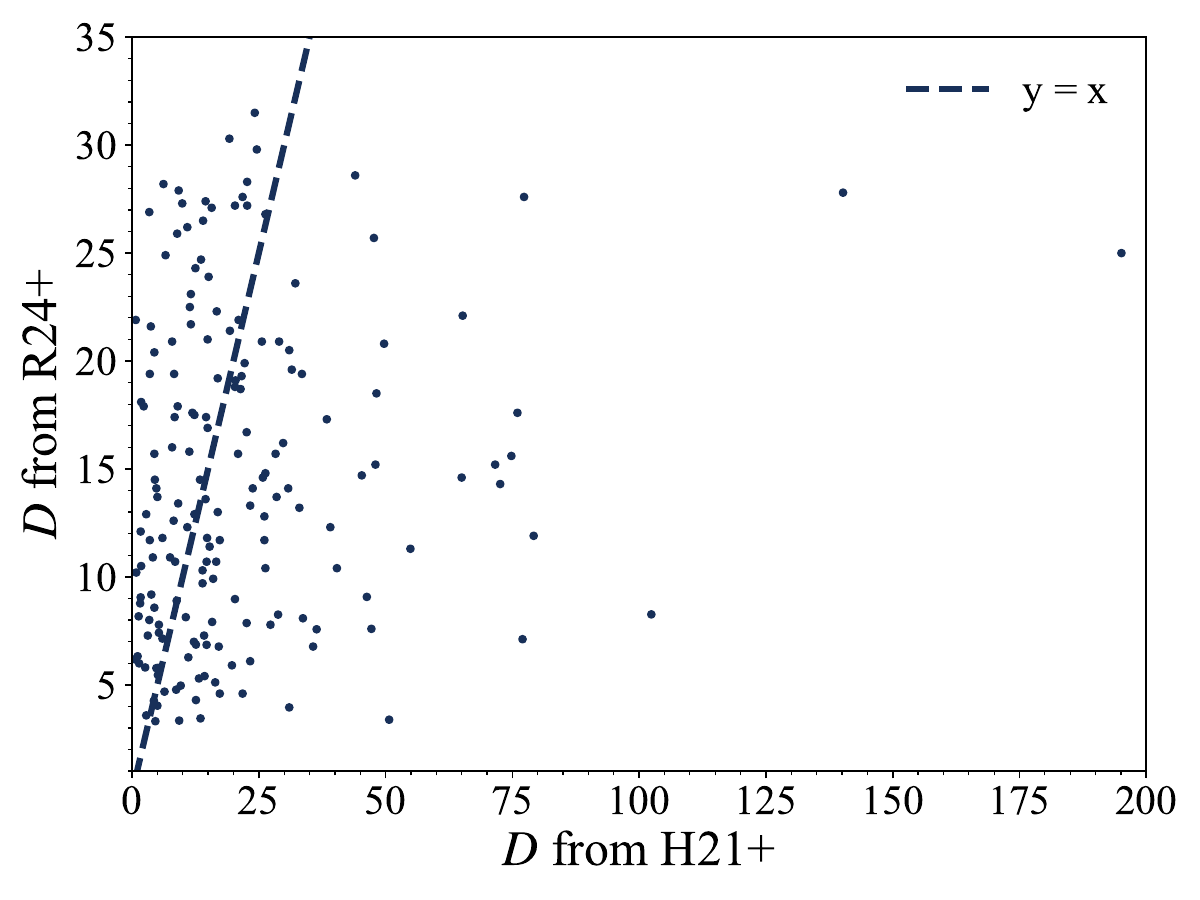}
        \caption{Scatter plot of Doppler factors $D$ among sources present both in H21+ and R24+ source lists.\label{fig:scatter}}
    \end{figure}
    \begin{figure}
        \centering
        \includegraphics[width=0.49\textwidth]{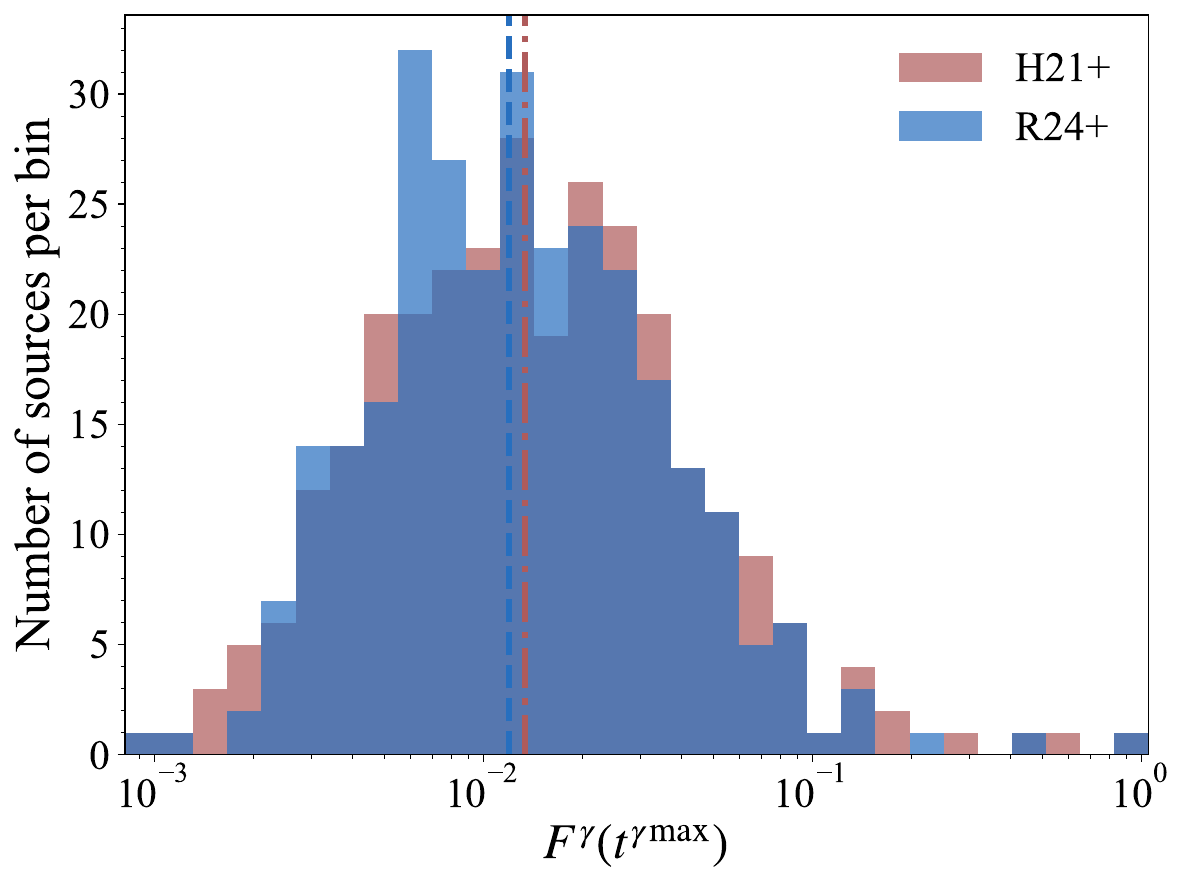}
        \caption{Same as Fig.~\ref{fig:redshift} but for normalised flux maxima $F^{\gamma}(t^{\gamma \, \max})$.\label{fig:fluxmax}}
    \end{figure}
    There are $174$ sources present in both H21+ and R24+ source lists as visualised in Fig.~\ref{fig:Venn}. The empirical distribution of source redshifts for both source lists is visualised as histograms in Fig.~\ref{fig:redshift}. Histograms of Doppler factors are plotted in Fig.~\ref{fig:doppler}, and a scatter plot of Doppler factors of sources present in both the H21+ and R24+ source lists is shown in Fig.~\ref{fig:scatter}. A histogram of normalised source flux maxima is shown in Fig.~\ref{fig:fluxmax}. While both H21+ and R24+ source lists have similar distributions of redshifts and global maxima of $\gamma$-ray fluxes, the Doppler factors have a substantial difference, as can be seen in Fig.~\ref{fig:scatter}. We discuss implications if this difference in Sect.~\ref{sec:discussion}.

\section{Methods}\label{sec:methods}
    In this section, we describe our procedure for searching for correlations between blazars in the given blazar list and neutrino alerts from IceCat-1. Our results follow in Sect.~\ref{sec:results}.

    Before doing any statistical test, we perform a spatial association of the neutrinos (both real $\{ x_{\nu} \}$ and mock $\{ m^{f}_{\nu} \}$) with the sources in the given blazar list (for each of the two tested source lists H21+ and R24+ separately): blazar $b$ is considered to be associated with neutrino $\nu$ if
    \begin{equation}
        \left( \frac{d^{\alpha}_{\nu b}}{\tilde{\alpha}_{\nu}^{\pm}} \right)^{2} + \left( \frac{d_{\nu b}^{\delta}}{\tilde{\delta}_{\nu}^{\pm}} \right)^{2} \leq 1,
        \label{eq:max_separation}
    \end{equation}
    where
    \begin{equation}
        \tilde{\alpha}_{\nu}^{\pm} = \alpha_{\nu}^{\pm} + \Delta,
        \label{eq:alpha_Delta}
    \end{equation}
    \begin{equation}
        \tilde{\delta}_{\nu}^{\pm} = \delta_{\nu}^{\pm} + \Delta,
        \label{eq:delta_Delta}
    \end{equation}
    where $d^{\alpha}_{\nu b}$ is the RA of blazar $b$ minus the RA of neutrino $\nu$, $d^{\delta}_{\nu b}$ is the declination of blazar $b$ minus the declination of neutrino $\nu$, and the sign $+$ or $-$ in $\alpha_{\nu}^{\pm}$ and $\delta_{\nu}^{\pm}$ is chosen according to the sign of $d^{\alpha}_{\nu b}$ and $d^{\delta}_{\nu b}$; $\Delta = 0.78^{\circ}$ is the assumed systematic angular uncertainty. This value was found \citep{Plavin:2022oyy} to give the optimal correlation strength between the radio-bright blazars and neutrino alerts, and we keep it fixed to avoid statistical penalising for multiple trials. A similar value of $1^{\circ}$ was found \citep{Kouch:2024xtd} to be optimal in the search for spatio-temporal correlations between the multiwavelength light curves of the CGRaBS blazars \citep{Healey:2007gb} and the \textit{IceCube} high-energy neutrinos. In a dedicated study with synthesised catalogues and a controlled number of synthetic blazar-neutrino correlations, \citet{Kouch:2025dgz} showed that enlarging the value of $\Delta$ does not lead to an increase in chances of finding a false-positive statistically significant correlation.

    Following Eq.~(1) in the work by \citet{Kouch:2025dgz}, we define for each of the alerts its enlarged angular uncertainty area as
    \begin{equation}
        U_{\nu} = \frac{\pi}{4} \left(\tilde{\alpha}^{+}_{\nu}\tilde{\delta}^{+}_{\nu} + \tilde{\alpha}^{-}_{\nu}\tilde{\delta}^{+}_{\nu} + \tilde{\alpha}^{-}_{\nu}\tilde{\delta}^{-}_{\nu} + \tilde{\alpha}^{+}_{\nu}\tilde{\delta}^{-}_{\nu}\right),
        \label{eq:uncertainty_area}
    \end{equation}
    which is used in our tests in factor $\rho_{\nu}$ penalising alerts with a large arrival uncertainty.

    Approximate linearity between $\gamma$-ray and neutrino fluxes may be expected in scenarios when the target photon fields for $p\gamma$ interactions are produced either internally (synchrotron or synchrotron-self-Compton photons) in the jet rest frame or externally in the host galaxy rest frame (external photons, e.g., from the radiation of the accretion disc, broad line region, or dusty torus). According to \citet{Dermer:2012rg} (see their Sect. 7), the Doppler amplification $\propto D^{p}$ makes both the observed neutrino and $\gamma$-ray fluxes boosted with $p = 4$ if neutrinos are produced in $p\gamma$ interactions on internal photon fields (of synchrotron origin) co-moving with the jet frame. However, $p \geq 5$ is expected if $p\gamma$ interactions are dominated by the photon fields external to the relativistic jet, with especially strong amplifications $p > 5$ if $p\gamma$ photopion production occurs near the threshold. The $\gamma$-ray external Compton emission is, in turn, produced with amplification $\propto D^{6}$. Thus, depending on a particular scenario, neutrino and gamma-ray fluxes may be related as $F_{\nu} \propto F_{\gamma}$ (internal photons of synchrotron origin) or $F_{\nu} \propto F_{\gamma}^{\geq 5/6}$ (external photons). If the observed $\gamma$ rays are of hadronic origin, i.e. produced not by primary electrons but originate from the same $p\gamma$ collisions as neutrinos and are a result of a hadron-initiated electromagnetic cascade, then neutrino and $\gamma$-ray fluxes can be expected to be linearly proportional because they originate from decays of unstable particles (mostly pions) produced in the same initial $p\gamma$ reactions. These considerations justify our assumption of the linear scaling between $\gamma$-ray and neutrino fluxes. \citet{Franckowiak:2020qrq} found that neutrino-emitting blazar candidates are statistically compatible with hypotheses of both a linear correlation and no correlation between neutrino and gamma-ray energy fluxes. Given that all the sources have measured redshifts and Doppler factors, using intrinsic luminosity in our study is possible as well, but is not warranted, since the expected number of neutrinos is proportional to the observed neutrino flux rather than the intrinsic neutrino luminosity; the recent work by \cite{Luo:2026nak} also showed that in searches for correlations between \textit{IceCube} neutrinos and X-ray-bright AGNs a robust correlation of neutrino and X-ray fluxes may indicate a genuine physical link while a luminosity-luminosity correlation alone is insufficient to establish a physical relationship between neutrino and X-ray emission in AGNs.

    We expect the jet-frame time delay $t^{\prime}_{\mathrm{delay}}$ to relate to the photopion proton energy-loss timescale $t^{\prime}_{p\gamma}$ and/or proton acceleration timescale $t^{\prime}_{\mathrm{acc}}$. They can vary significantly depending on the proton energy and parameters of photon fields serving as the target for the photopion production process, on the magnetic field, and proton acceleration mechanism. A priori, we do not know the exact value of $t^{\prime}_{\mathrm{delay}}$. Thus, we perform a scan over parameter $t^{\prime}_{\mathrm{delay}} \sim U [t^{\prime}_{\min}; t^{\prime}_{\max}]$, $t^{\prime}_{\min} = \Delta_{F} = 1$~month corresponding to the bin size $\Delta_{F}$ of the \textit{Fermi}-LAT light curves. \citet{Podlesnyi:2025aqb}, in modelling the brightest \textit{Fermi}-LAT blazar flare of 3C 454.3, obtained that close to the threshold of the photopion production around $E_{\nu} \sim 100$~TeV, $t^{\prime}_{\mathrm{delay}} \approx  t^{\prime}_{p\gamma} \approx t^{\prime}_{\mathrm{acc}} \sim 10^{9}$~s (see Fig.~12 by \citealp{Podlesnyi:2025aqb}). Thus, we adopt $t^{\prime}_{\max} = 3 \times 10^{9}$~s $= 3.47\times10^4$~d being agnostic of the exact mechanism of the possible $\nu$ delay and no particular preference to any delay between $t^{\prime}_{\min}$ and $t^{\prime}_{\max}$. The chosen value of $t^{\prime}_{\max}$, from the perspective of observable quantities, is approximately equal to the median of maximum delays $[(t_{\nu}^{\max}- t_{\mathrm{LAT}}^{\min}) D_{\mathrm{median}} / (1 + z_{\mathrm{median}})]$, where $t_{\nu}^{\mathrm{max}} = 60231.917$~(MJD) is the time of the last alert in IceCat-1, $t_{\mathrm{LAT}}^{\min} = 54\,698$~(MJD) is the earliest time in the \textit{Fermi}-LAT light curves, and the median values of redshifts $z_{\mathrm{median}}$ and Doppler factors $D_{\mathrm{median}}$ are shown as vertical lines in Figs.~\ref{fig:redshift} and~\ref{fig:doppler} respectively. The time step in our scan approximately equals $29$ days (corresponding to $1~200$ steps in $t^{\prime}_{\mathrm{delay}}$) so that a hypothetical source with $D = 1$ and $z = 1$ does not skip a single \textit{Fermi}-LAT time bin given the $\pm 1$~month margin we use in our analysis as explained below.
    
    For each alert $\nu$, which has at least one blazar from the given source list in its uncertainty region (hereafter ``associated alert''), we assign a weight
    \begin{multline}
        w_{\nu}(t^{\prime}_{\mathrm{delay}}) = \\ = \sum_{b} \left[s_{\nu} \rho_{\nu} G_{\nu b}(t^{\gamma \max}_{b}, t_{\nu}; t^{\prime}_{\mathrm{\, delay}}, D_{b}, z_{b})F^{\gamma}_{b}(t^{\gamma \max}_{b}) \right],
    \label{eq:weights}
    \end{multline}
    where the sum $\sum_{b}$ runs over all blazars within the uncertainty region of neutrino $\nu$ (as was pointed out by \citet{Kouch:2025ipk}, a single neutrino cannot come from more than one source, but having several objects in its angular uncertainty region increases the probability of this neutrino coming from a blazar), $s_{\nu}$ is the alert signalness, $\rho_{\nu}$ is the factor penalising alerts with too large uncertainties. Following recommendations of \citet{Kouch:2025dgz}, we use a top-hat spatial factor penalising alerts with too large angular uncertainties:
    \begin{equation}
        \rho_{\nu} = \begin{cases}
            1, \mathrm{\, if \,} U_{\nu} \leq \tilde{U}_{\nu}\\
            \displaystyle{\left(\frac{\tilde{U}_{\nu}}{U_{\nu}}\right)}, \mathrm{\, if \,} U_{\nu} > \tilde{U}_{\nu},
        \end{cases}
    \end{equation}
    where $\tilde{U}_{\nu} = 15.6$~sq.~deg. is the median angular uncertainty (see Eq. \ref{eq:uncertainty_area}) of IceCat-1 alerts (after the extension by $\Delta = 0.78^{\circ}$).
    
    The key point of the current paper is the search for neutrino-blazar correlations taking into account the possibility of neutrino emission lagging behind the $\gamma$-ray flare due to slowness of the $p\gamma$ photopion production and/or proton acceleration as was suggested by \citet{Podlesnyi:2025aqb}. Given that even during the brightest blazar flares, the \textit{IceCube} neutrino yield is of the order of $10^{-3}-10^{-2}$ muon neutrinos at energies $\gtrsim 100$~TeV \citep{Oikonomou:2019djc, Kreter:2020kpm, Robinson:2024hmz} and during quieter activity periods even lower (assuming proportionality between $\gamma$-ray and $\nu$ fluxes), we assume that if the neutrino is indeed associated with the blazar, it must have come in a causal connection with the major flare. Thus, the last term $F^{\gamma}_{b}(t^{\gamma \max}_{b})$ in Eq.~(\ref{eq:weights}) denotes the global maximum of the \textit{Fermi}-LAT light curve of blazar $b$ detected at $t^{\gamma \max}_{b}$ among the times \textit{prior to} the associated neutrino arrival ($+\Delta_{F} = 1$~month to account for the size of the time bin and still allow for instantaneous associations) normalized by $1.66 \times 10^{-5}$~cm$^{-2}$~s$^{-1}$ (the maximum value of monthly-averaged fluxes among the studied blazars within the whole \textit{Fermi}-LAT observation history observed from FSRQ 3C 454.3 during the brightest \textit{Fermi}-LAT blazar flare in November 2010; see, e.g., \citealp{Fermi-LAT:2011iez,Nalewajko:2012yf,Podlesnyi:2025aqb}). As for the temporal term $G_{\nu b}$ accounting for the neutrino delay, we define it by linking to the time $t^{\gamma \max}_{b}$ of the global maximum of the associated blazar light curve \textit{prior to} the neutrino arrival as follows
    \begin{multline}
        G_{\nu b}(t^{\gamma \max}_{b}, t_{\nu}; t^{\prime}_{\mathrm{\, delay}}, D_{b}, z_{b}) =\\
        =\begin{cases}
        0, \text{\, if } t^{\gamma \max}_{b} > t_{\nu} + \Delta_{F}, \\
        1, \text{\, if } \bigl|\;| t^{\gamma \max}_{b} - t_{\nu}| - t_{\text{b delay}}\;\bigr| \leq \Delta_{F}, \\
        \exp\!\left(
            -\dfrac{\bigl(t^{\gamma \max}_{b} - t_{\nu} + t_{\text{b delay}}\bigr)^2}
                   {2(t_{\text{b delay}})^2}
        \right), \text{\, otherwise.}
        \end{cases}
        \label{eq:lagging_weight}
    \end{multline}
    where the delay for source $b$ in the observer's frame is given by
    \begin{equation}
        t_{b\mathrm{\, delay}}(t^{\prime}_{\mathrm{delay}}, D_{b}, z_{b}) = \max \left\{\Delta_{F}, \frac{1 + z_{b}}{D_{b}} t^{\prime}_{\mathrm{delay}} \right\}.
    \label{eq:delay_Earth_frame}
    \end{equation}

    \begin{figure}
        \centering
        \includegraphics[width=0.45\textwidth]{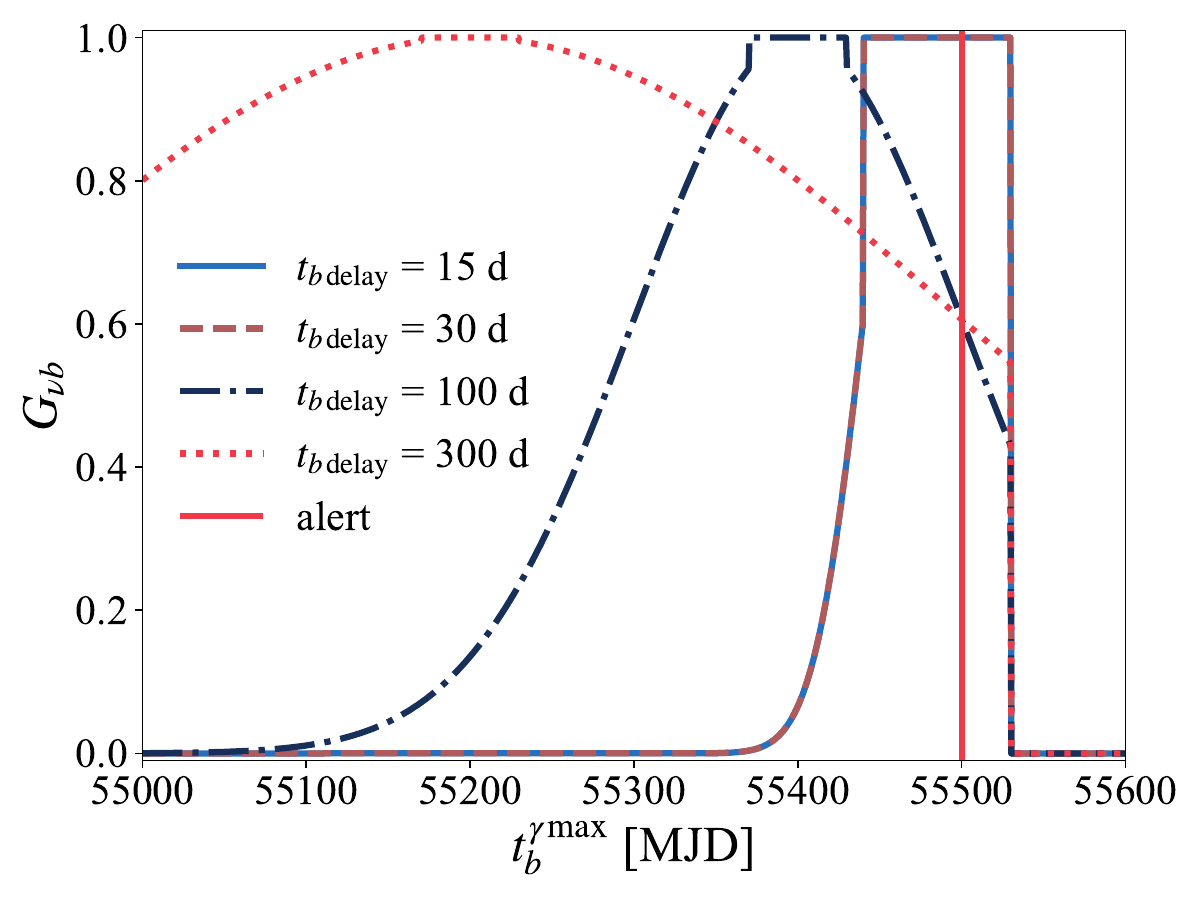}
        \caption{An example of the graph of the temporal part $G_{\nu b}$ of weights defined in Eq. (\ref{eq:lagging_weight}) as a function of $t^{\gamma \max}_{b}$ for a fixed $t_{\nu}$ (shown as the red solid vertical line) and a set of fixed $t_{b\mathrm{\, delay}}$ (see the legend).\label{fig:G}}
    \end{figure}
    
    In Eq. (\ref{eq:lagging_weight}) in the first row we, for the sake of completeness, set $G_{\nu b}$ as zero for $t^{\gamma \max}_{b} > t_{\nu} + \Delta_{F}$, albeit technically $G_{\nu b}$ is not defined for these times because by definition $t^{\gamma \max}_{b}$ is the time of the maximum of the \textit{Fermi}-LAT light curve of blazar $b$ only among times $\leq t_{\nu} + \Delta_{F}$. The second row in Eq. (\ref{eq:lagging_weight}) assigns equal weights to all values of $t^{\gamma \max}_{b}$ if they preceed $t_{\nu}$ by $t_{b\mathrm{\, delay}}$ with an absolute margin of $\Delta_{F}$ to account for the finite size of the \textit{Fermi}-LAT time bin. The third row handles all other cases, assigning the Gaussian penalty with the width equal to the anticipated neutrino delay $t_{b\mathrm{\, delay}}$ in the Earth frame. The illustration of the graph of $G_{\nu b}$ as a function of $t^{\gamma \max}_{b}$ for a fixed $t_{\nu}$ and $t_{b\mathrm{\, delay}}$ is given in Fig.~\ref{fig:G}.

    The global test statistic (TS) for the whole dataset is just the sum of individual weights:
    \begin{equation}
        \mathrm{TS}(t^{\prime}_{\mathrm{delay}}) = \sum\limits_{\nu} w_{\nu}(t^{\prime}_{\mathrm{delay}}),
        \label{eq:TS}
    \end{equation}
    and TS is a function of the jet-frame neutrino time delay $t^{\prime}_{\mathrm{delay}}$, assumed to be universal among all sources.

    We compare the obtained $\mathrm{TS}$ value for the real dataset $\{ x_{\nu} \}$ against the distribution of values $\mathrm{TS}_{f}$ obtained for the set of mock datasets $\{ m_{\nu}^{f} \}$ with values of right ascensions $\alpha_{\nu}^{f}$ and arrival times $t_{\nu}^{f}$ randomly generated as described in Sect.~\ref{sec:mock}. Then, for each correlation search (test) performed, we, following \citet{Plavin:2020emb,Davison_Hinkley_1997}, obtain the p-value
    \begin{equation}
        p(t^{\prime}_{\mathrm{delay}}) = \frac{\# \{\mathrm{TS}_{f} \geq \mathrm{TS} \}_{f = 1}^{N_f} + 1}{N_{f} + 1},
        \label{eq:p_value}
    \end{equation}
    i.e. $p(t^{\prime}_{\mathrm{delay}})$ is the fraction of times the test statistic for a mock dataset $\mathrm{TS}_{f}^{i}$ was larger than or equal to the TS value for the real dataset $\{ x_{\nu} \}$ at the given value of the $\nu$ time delay $t^{\prime}_{\mathrm{delay}}$.

\section{Results}\label{sec:results}
    
\begin{figure}[t]
        \centering
        \includegraphics[width=0.45\textwidth]{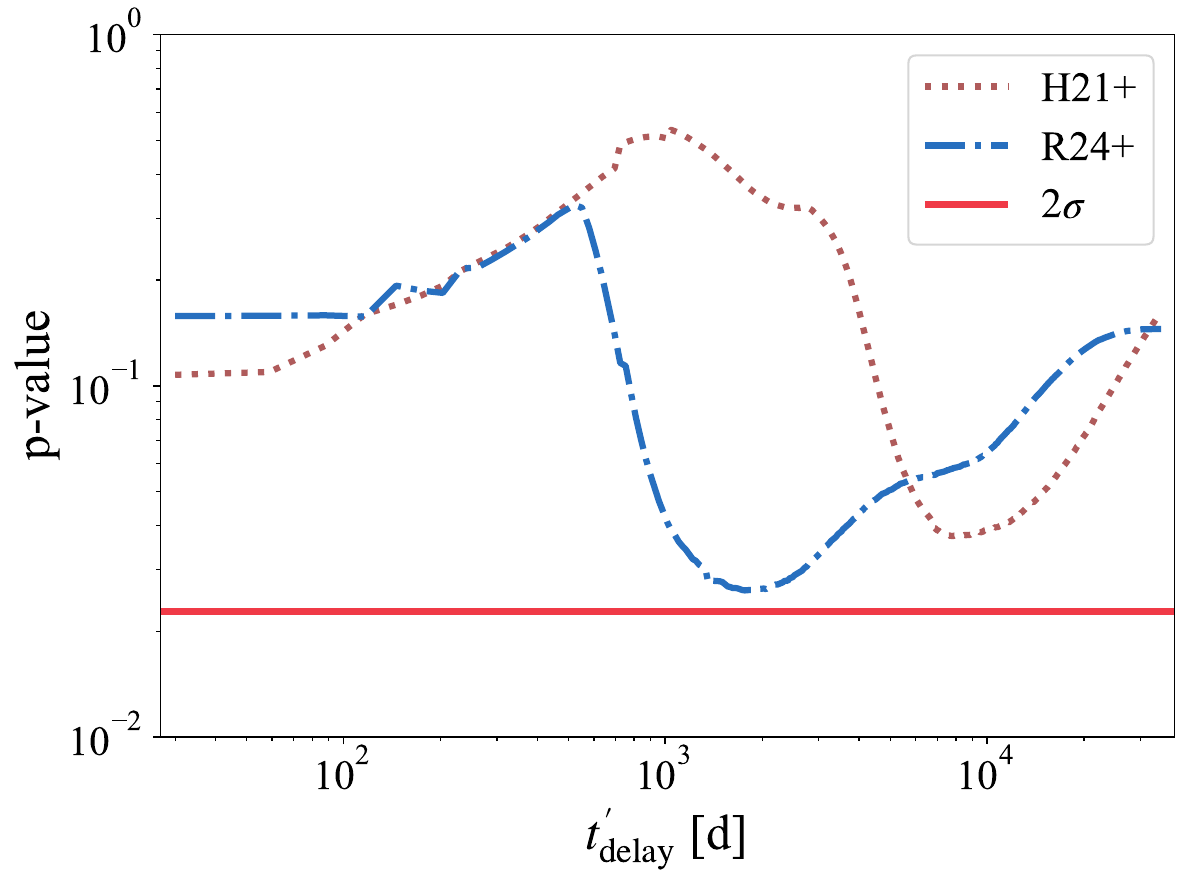}
        \caption{$p$-value (see Eq.~\ref{eq:p_value}) as a function of the jet-frame time delay assumed to be universal for all AGNs for the test with the sources from H21+ (dotted curve) and sources from R24+ (dashed-dotted curve).\label{fig:p_value_plot}}
    \end{figure}

    The results of the tests performed for H21+ and R24+ source lists are shown in Fig.~\ref{fig:p_value_plot}. For the H21+ source list, the global minimum of p-value is $0.037$ reached at the jet-frame delay $\tilde{t}^{\prime}_{\mathrm{delay}} = 7.7 \times 10^{3}$~d. For the R24+ source list, the minimum of $p(\hat{t}^{\prime}_{\mathrm{delay}}) = 0.026$ is reached at $\hat{t}^{\prime}_{\mathrm{delay}} = 1.9 \times 10^{3}$~d. Tables \ref{tab:top_ten_Homan} and \ref{tab:top_ten_Rodrigues} present the top ten associations with the highest values of $\mathrm{TS}_{\nu}$ for H21+ and R24+, respectively.

    In Fig.~\ref{fig:p_value_plot}, one can see that the p-value curve of R24+ almost reaches the $2\sigma$ level (one-tailed, see, e.g., Eq.~(40.46) of \citealp{ParticleDataGroup:2020ssz}). The trial correction (for multiple tested values of $t^{\prime}_{\mathrm{delay}}$) performed according to the prescriptions of \citet{Plavin:2020emb} shows, however, that the minima of the pre-trial p-values as deep as in Fig.~\ref{fig:p_value_plot} occur by chance in $p_{\mathrm{post-trial\, H21+}} = 14$\% ($1.06\sigma$) and $p_{\mathrm{post-trial\, R24+}} = 11$\% ($1.21\sigma$) among mock datasets for the tests with H21+ and R24+ source lists respectively. Thus, none of the obtained p-values shows evidence of a significant association between IceCat-1 neutrinos and AGNs from the considered source lists.

    We note that the optimal delay $\tilde{t}^{\prime}_{\mathrm{delay}} = 7.7 \times 10^{3}$~d for H21+ is by $4$ times greater than $\hat{t}^{\prime}_{\mathrm{delay}} = 1.9 \times 10^{3}$~d for R24+. This is driven by the substantial difference in the Doppler factors (compare Tables~\ref{tab:top_ten_Homan} and \ref{tab:top_ten_Rodrigues}) of 3C~454.3 and 3C~279, that contribute $53$\% of the total TS for the former and $80$\% of the total TS for the latter: \citet{Homan:2021ijm} estimates the Doppler factor of 3C~279 by a factor of five larger than the estimate of \citet{Rodrigues:2023vbv}, and the Doppler factor of 3C~454.3 by a factor of three larger than that found by \citet{Rodrigues:2023vbv}.
    
\begin{figure*}[t]
        \centering
        \includegraphics[width=0.80\textwidth]{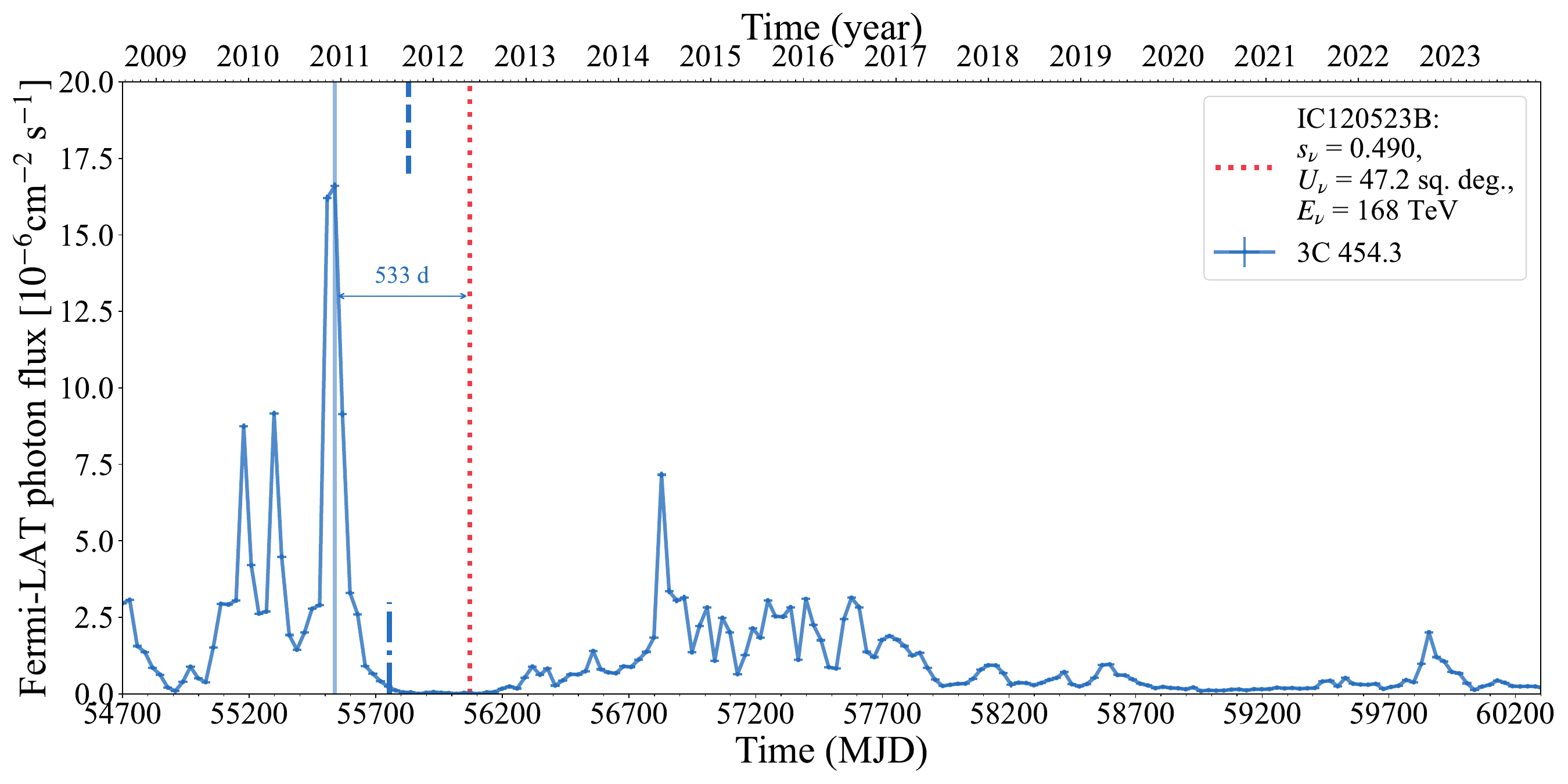}
        \\
        \includegraphics[width=0.80\textwidth]{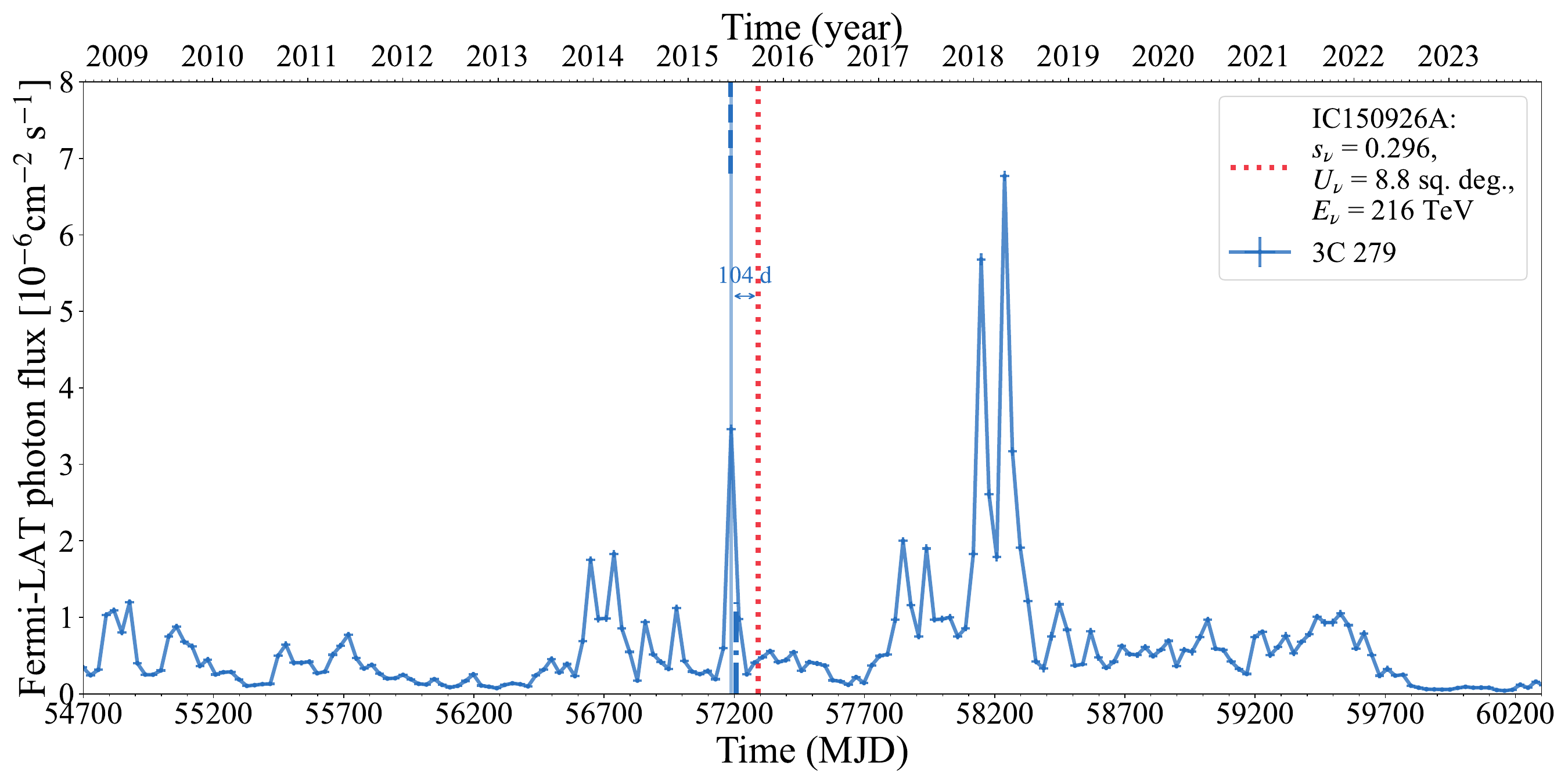}
        \caption{Light curves of 3C 454.3 (upper panel), and 3C 279 (lower panel), which are among the most significant associations between IceCat-1 neutrinos and AGNs in both the tests (see Tables~\ref{tab:top_ten_Homan} and \ref{tab:top_ten_Rodrigues}). The connected markers with error bars show monthly-binned \textit{Fermi}-LAT photon fluxes in the energy range 100~MeV--100~GeV from the public repository \citep{Fermi-LAT:2023iml} (ordinate axis). Dotted vertical lines indicate the time $t_{\nu}$ of the neutrino arrival. Time intervals between the alert and the global maximum of the light curve prior to $t_{\nu}$ are shown as double-pointed arrows. Vertical upper dashed risks indicate the anticipated position of the maximum with the Doppler factor from R24+ and $\hat{t}^{\prime}_{\mathrm{delay}} = 1.9 \times 10^{3}$~d, and vertical lower dashed-dotted risks --- with the Doppler factor from H21+ and $\tilde{t}^{\prime}_{\mathrm{delay}} = 7.7 \times 10^{3}$~d. The width of the Gaussian temporal factor $G_{\nu b}$ in weights is equal to the distance between the vertical risks and the line of $t_{\nu}$ (see Eq. \ref{eq:lagging_weight}).\label{fig:light_curves_I}}
    \end{figure*}

\begin{figure*}[t]
        \centering
        \includegraphics[width=0.80\textwidth]{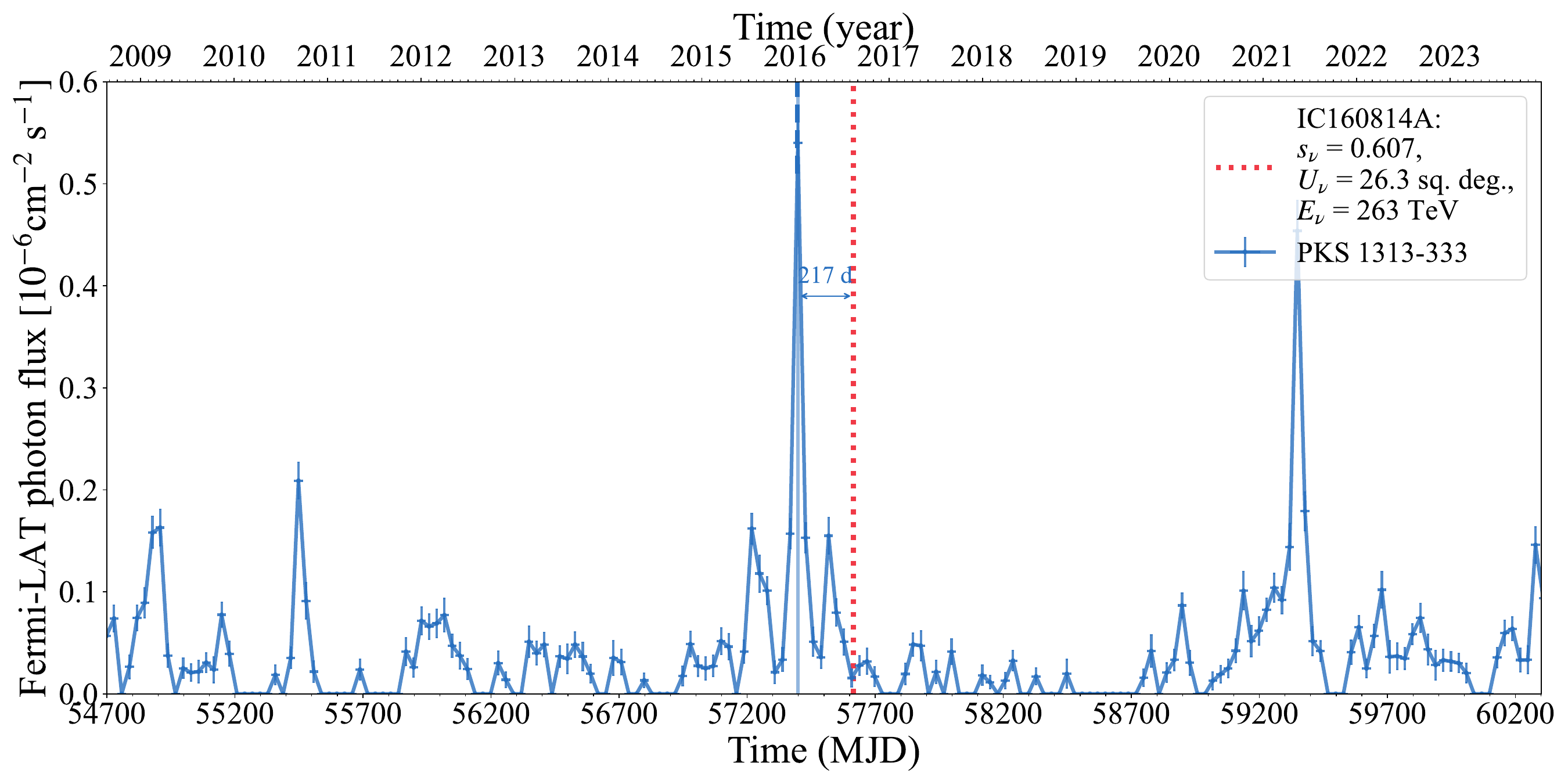}
        \\
        \includegraphics[width=0.80\textwidth]{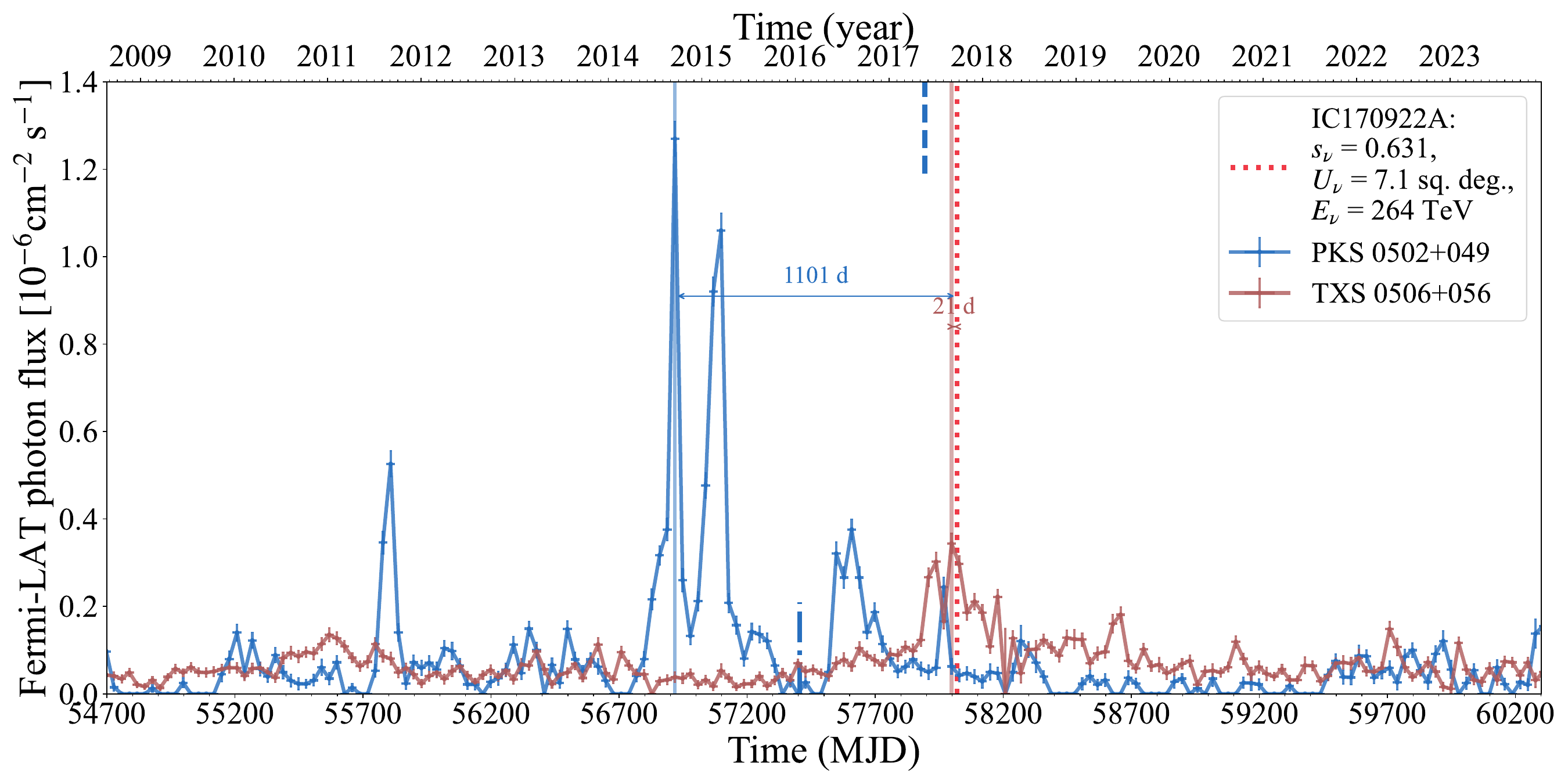}
        \caption{Same as Fig.~\ref{fig:light_curves_I}, but for PKS~1313-333 and PKS~0502+049 / TXS~0506+056.\label{fig:light_curves_II}}
    \end{figure*}

    We show examples of two blazar light curves for associated alerts with the two highest values of $\mathrm{TS}_{\nu}$ in both of the two tests performed in Fig.~\ref{fig:light_curves_I}, and the light curves of the associations with the third-highest $\mathrm{TS}_{\nu}$ in Fig.~\ref{fig:light_curves_II}.

    The most significant association is driven by the alert IC120523B\footnote{In the original publication of \citet{IceCube:2023htm}, this alert had a name IC120523A, which was also given to another alert in the catalogue; in the updated version of IceCat-1 \citep{IceCat-1v4}, the alert has the name IC120523B.} associated with the brightest \textit{Fermi}-LAT blazar flare of the FSRQ 3C 454.3, which happened in November 2010 \citep{Fermi-LAT:2011iez,Nalewajko:2012yf,Podlesnyi:2025aqb}. Despite large angular uncertainty of $47.2$ sq. deg., the exceptional brightness of 3C 454.3 during that flare and that the neutrino arrival delay in the Earth frame is not far from the anticipated $(t^{\prime}_{\mathrm{delay}} (1 + z) / D)$ for the both tested source lists, provides the highest value of $\mathrm{TS}_{\nu}$ in both Tables \ref{tab:top_ten_Homan} and \ref{tab:top_ten_Rodrigues}. 3C 454.3 is well within the original \textit{IceCube} angular uncertainty contour lying in $0.73^{\circ}$ from the best-fit arrival direction of IC120523B, and this association does not require its extension by $\Delta$.

    3C 279 is the second brightest FSRQ in the \textit{Fermi}-LAT 4FGL catalogue \citep{Fermi-LAT:2019yla}. The neutrino IC150926A arrives around $100$~d (mind the finite size of one month of the \textit{Fermi}-LAT light curves) after the bright but short multiwavelength flare of 3C 279 happened around MJD 57~200 detected as well in X-rays and optical photons \citep{Paliya:2015tea,Pittori:2018hmw,Singh:2020upf}. The peak of the flare is in perfect agreement with the anticipated neutrino lagging by $(t^{\prime}_{\mathrm{delay}} (1 + z) / D)$ for both H21+ and R24+ source lists. As we mentioned above, the four-times difference in estimates of Doppler factors by \citet{Homan:2021ijm} and \citet{Robinson:2024hmz} seems to drive the difference in the optimal delays providing the minima of p-values in Fig.~\ref{fig:p_value_plot}. The combination of the source brightness and the prior $\gamma$-flux global maximum happening at the anticipated time, together with a relatively small angular uncertainty of $8.8$~sq. deg., result in the second highest value of $\mathrm{TS}_{\nu}$ in both Tables~\ref{tab:top_ten_Homan} and \ref{tab:top_ten_Rodrigues}. This association requires the extension of the original \textit{IceCube} angular uncertainty contour by $\Delta = 0.78^{\circ}$.

    \begingroup 
    \setlength{\tabcolsep}{6pt} 
    \setlength\extrarowheight{2pt}
    \begin{table*}
        \centering
        \begin{tabular}{|l|l|c|c|c|c|c|c|c|c|c|}
        \hline
        Alert & Association & $s_{\nu}$ & $E_{\nu}$ [TeV] & $U_{\nu}$ [$^{{\circ}^2}$] & $t_{\nu}$ & $t^{\gamma \max}_{b}$ & $F^{\gamma}_{b}(t^{\gamma \max}_{b})$ & $\mathrm{\mathrm{TS}_{\nu}}$ & $D_{b}$ & $z_{b}$\\
        \hline
        IC120523B & 3C 454.3 & 0.490 & 168 & 47.2 & 56071 & 55538 & 1.000 & 0.1287 & 45.3 & 0.86\\
        \hline
        IC150926A & 3C 279 $\dagger$ & 0.296 & 216 & 8.8 & 57292 & 57188 & 0.208 & 0.0617 & 140.2 & 0.54\\
        \hline
        IC170922A & PKS 0502+049 $\dagger$ & 0.631 & 264 & 7.1 & 58019 & 56918 & 0.077 & 0.0432 & 24.6 & 0.95\\
         & TXS 0506+056 &  &  &  &  & 57998 & 0.021 &  & 1.8 & 0.34\\
        \hline
        IC221223A & B2 2308+34 $\dagger$ & 0.795 & 353 & 7.5 & 59936 & 59588 & 0.028 & 0.0181 & 22.8 & 1.82\\
        \hline
        IC180608A & PKS 0440-00 $\dagger$ & 0.396 & 158 & 11.3 & 58278 & 56498 & 0.047 & 0.0168 & 3.5 & 0.45\\
        \hline
        IC160727A & 4C +14.23 $\dagger$ & 0.296 & 105 & 14.9 & 57596 & 55148 & 0.055 & 0.0082 & 13.9 & 1.04\\
        \hline
        IC181023B & TXS 0518+211 $\dagger$ & 0.427 & 136 & 11.5 & 58415 & 56588 & 0.018 & 0.0071 & 2.7 & 0.11\\
        \hline
        IC131014A & MG1 J021114+1051 $\dagger$ & 0.665 & 293 & 6.3 & 56580 & 55568 & 0.011 & 0.0071 & 6.3 & 0.20\\
        \hline
        IC190410A & PKS 2029+121 & 0.280 & 105 & 44.6 & 58583 & 54848 & 0.004 & 0.0050 & 28.3 & 1.22\\
         & PKS 2032+107 &  &  &  &  & 57278 & 0.051 &  & 9.7 & 0.60\\
        \hline
        IC120515A & OP 313 & 0.613 & 194 & 13.3 & 56063 & 54698 & 0.012 & 0.0039 & 24.2 & 1.00\\
        \hline
        \end{tabular}
        \caption{Top ten neutrino alerts with the highest values of their $\mathrm{TS}_{\nu}$ in the test with the source list H21+ corresponding to the local minimum of $p(\tilde{t}^{\prime}_{\mathrm{delay}}) = 0.037$ reached at $\tilde{t}^{\prime}_{\mathrm{delay}} = 7.7 \times 10^{3}$~d. The sum $\mathrm{TS}$ for the whole dataset is $0.3577$. The symbol $\dagger$ indicates associations requiring the extension of the original \textit{IceCube} errors by $\Delta = 0.78^{\circ}$. Times $t_{\nu}$ and $t^{\gamma \max}_{b}$ of the neutrino arrival and of the maximum of the light curve of the associated source $b$ among the times before $t_{\nu}$ are in MJD.\label{tab:top_ten_Homan}}
    \end{table*}
    \begin{table*}
    \setlength{\tabcolsep}{6pt} 
    \setlength\extrarowheight{2pt}
        \centering
        \begin{tabular}{|l|l|c|c|c|c|c|c|c|c|c|}
        \hline
        Alert & Association & $s_{\nu}$ & $E_{\nu}$ [TeV] & $U_{\nu}$ [$^{{\circ}^2}$] & $t_{\nu}$ & $t^{\gamma \max}_{b}$ & $F^{\gamma}_{b}(t^{\gamma \max}_{b})$ & $\mathrm{\mathrm{TS}_{\nu}}$ & $D_{b}$ & $z_{b}$\\
        \hline
        IC120523B & 3C 454.3 & 0.490 & 168 & 47.2 & 56071 & 55538 & 1.000 & 0.0787 & 14.7 & 0.86\\
        \hline
        IC150926A & 3C 279 $\dagger$ & 0.296 & 216 & 8.8 & 57292 & 57188 & 0.208 & 0.0617 & 27.8 & 0.54\\
        \hline
        IC160814A & PKS 1313-333 $\dagger$ & 0.607 & 263 & 26.3 & 57615 & 57398 & 0.033 & 0.0117 & 19.3 & 1.21\\
        \hline
        IC211208A & PKS 0735+17 $\dagger$ & 0.502 & 171 & 25.1 & 59557 & 59558 & 0.019 & 0.0036 & 5.8 & 0.42\\
        \hline
        IC150812B & PKS 2145+06 $\dagger$ & 0.831 & 508 & 6.5 & 57247 & 56648 & 0.003 & 0.0025 & 4.0 & 1.00\\
        \hline
        IC140927A & PKS 0336-01 $\dagger$ & 0.481 & 182 & 37.2 & 56927 & 56948 & 0.018 & 0.0019 & 26.8 & 0.85\\
        \hline
        IC170427A & RX J0011.5+0058 $\dagger$ & 0.383 & 155 & 40.3 & 57870 & 57578 & 0.012 & 0.0018 & 17.3 & 1.49\\
         & S3 0013-00 &  &  &  &  & 56018 & 0.008 &  & 15.0 & 1.58\\
        \hline
        IC120605A & OL 318 $\dagger$ & 0.385 & 107 & 21.1 & 56084 & 55868 & 0.003 & 0.0016 & 19.9 & 1.41\\
         & B2 1015+35B $\dagger$ &  &  &  &  & 55148 & 0.003 &  & 3.6 & 1.23\\
        \hline
        IC200614A & B2 0202+31 & 0.415 & 115 & 65.1 & 59015 & 58598 & 0.021 & 0.0015 & 20.5 & 1.47\\
        \hline
        IC220225A & PKS 0215+015 $\dagger$ & 0.378 & 154 & 27.2 & 59636 & 59648 & 0.012 & 0.0015 & 14.6 & 1.72\\
        \hline
        \end{tabular}
        \caption{Top ten neutrino alerts with the highest values of their $\mathrm{TS}_{\nu}$ in the test with the source list R24+ corresponding to the global minimum of $p(\hat{t}^{\prime}_{\mathrm{delay}}) = 0.026$ reached at $\hat{t}^{\prime}_{\mathrm{delay}} = 1.9 \times 10^{3}$~d. The sum $\mathrm{TS}$ for the whole dataset is $0.1734$. Note the change of values for $D$ w.r.t. Table~\ref{tab:top_ten_Homan} and that TXS 0506+056 is absent in R24+. \label{tab:top_ten_Rodrigues}}
    \end{table*}
    \endgroup
    

\section{Caveats and additional tests}\label{sec:caveats}
    The statistical weight $w_{\nu}$ used in the analysis (see Eq.~\ref{eq:weights}) is proportional to the global maximum of the \textit{Fermi}-LAT light curve for a given object among the times \textit{prior} to the associated neutrino arrival. This choice has an important consequence: Most of the objects have their global maximum of $\gamma$-ray flux around two orders of magnitude below that of the brightest blazar flare in 3C~454.3, as can be easily seen in Fig.~\ref{fig:fluxmax}, and well-reconstructed events with high signalness $s_{\nu}$ associated with not so bright sources will be heavily penalised with respect to a dozen of brightest sources even if the neutrino associated with them may have a large angular uncertainty (as in the case for the IC120523B~--- 3C~454.3 association). This behaviour is anticipated. For instance, in the model of \citet{Podlesnyi:2025aqb}, 3C~454.3 alone is estimated to account for $\sim10$\% of observable \textit{IceCube} neutrino flux from all FSRQs at energies $\geq 100$~TeV, and the observable neutrino and gamma-ray fluxes are found to be in an approximate linear relation as implied in Eq.~(\ref{eq:weights}). This however, might be suboptimal from the perspective of searches for a statistical correlation: As demonstrated by \citet{Kouch:2025dgz} in their synthetic tests dedicated to finding the best strategy for searching correlations between neutrinos and electromagnetic sources, count-based global TS constructed as a sum of associations passing a certain threshold (say, in $\gamma$-ray flux or angular uncertainty area) performed, on average, better than an averaged-based TS, such as ours, where the associations with the brightest blazars drive the value of the global TS.
    
    To address this caveat, we perform an additional test where the last term $F^{\gamma}_{b}(t^{\gamma \max}_{b})$ of Eq.~(\ref{eq:weights}) is omitted, i.e. all associated events contribute to the global TS only based on $s_{\nu}$, $\rho_{\nu}$, and $G_{\nu b}$. This test gives null results with post-trial $p$ values higher than in our fiducial test described in Sect.~\ref{sec:methods}; its results are summarised in Tables~\ref{tab:top_ten_Homan_no_flux} and \ref{tab:top_ten_Rodrigues_no_flux} and in Fig.~\ref{fig:p_value_plot_no_flux} in the Appendix.

    Another caveat of our study is that statistical weights $w_{\nu}$ in Eq.~(\ref{eq:weights}) demand that a potential association of an \textit{IceCube} neutrino with an AGN is linked to a global maximum of the AGN light curve among the times prior to the neutrino arrival. However, blazar light curves generally show tens of prominent flares over many years, and each of those flares could be a proxy for neutrino emission. Our choice is dictated by two factors: (i) model predictions of \citet{Oikonomou:2019djc, Kreter:2020kpm, Robinson:2024hmz} showing that even during the brightest blazar flares, the \textit{IceCube} neutrino yield is of the order of $10^{-3}-10^{-2}$ muon neutrinos at energies $\gtrsim 100$~TeV, and a neutrino associated with a blazar has the highest probability to be associated with the brightest flare in that blazar; (ii) if a given neutrino-AGN association is genuine, trying multiple flares would inevitably increase statistical noise, thereby reducing the significance of such an association.
    
    Nevertheless, to check the effect of our choice linking the weights to the global maximum of an associated AGN light curve prior to the neutrino arrival, we conduct an additional test with weights defined as
    \begin{equation}
        \omega_{\nu}(t^{\prime}_{\mathrm{delay}}) = \sum_{b} \left[s_{\nu} \rho_{\nu} T_{\nu b}(t_{\nu}, t^{\prime}_{\mathrm{\, delay}}, D_{b}, z_{b}) \right],
    \label{eq:alternative_weights}
    \end{equation}
    where
    \begin{multline}
        T_{\nu b}(t_{\nu}, t^{\prime}_{\mathrm{\, delay}}, D_{b}, z_{b}) = \begin{cases}
            1, \mathrm{\, if \, }  F^{\gamma}_{b}(t_{\nu} - t_{b\mathrm{\, delay}}) \geq F^{68\%}_{b}, \\
            0, \mathrm{\, otherwise},
        \end{cases}
    \label{eq:alternative_temporal_weights}
    \end{multline}
    $t_{b\mathrm{\, delay}} = t_{b\mathrm{\, delay}}(D_{b}, z_{b})$ is the anticipated neutrino delay in the Earth rest frame defined in Eq.~(\ref{eq:delay_Earth_frame}), $F^{68\%}_{b}$ is the 68\% percentile of the $\gamma$-ray photon flux among the all observed \textit{Fermi}-LAT photon fluxes of the light curve of blazar $b$. If $t_{\nu} - t_{b\mathrm{\, delay}}$ falls before the start of the \textit{Fermi}-LAT mission, we treat the source flux at these times as zero. We allow a $\pm \Delta_{F}$ margin in Eq.~(\ref{eq:alternative_temporal_weights}) to account for the finite size of the time bins of the \textit{Fermi}-LAT light curves. Thus, the alternative temporal weight $T_{\nu b}$ in Eq.~(\ref{eq:alternative_temporal_weights}) is Boolean and returns unity if an associated blazar $b$ is in a flaring state defined here as a state with the $\gamma$-ray flux exceeding the $68$\% percentile of the $\gamma$-ray flux distribution for the given source. Thus, this Boolean definition of the temporal weight $T_{\nu b}$ does not require the connection of a given neutrino with only the brightest flare of the associated source and addresses the caveat of the dominance of the brightest sources discussed above. Our additional tests with alternative weights $\omega_{\nu}$ result in no significant correlations between IceCat-1 neutrinos and sources from H21+ and R24+ source lists with post-trial $p$ values higher than in our fiducial test described in Sect.~\ref{sec:methods}. Tables~\ref{tab:top_ten_Homan_68} and \ref{tab:top_ten_Rodrigues_68} with Fig.~\ref{fig:p_value_plot_68} in the Appendix summarise our results with this additional test.

    Lastly, motivated by \citet{Plavin:2020emb,Plavin:2022oyy}, that reported evidence of the correlation between the \textit{IceCube} neutrino alerts and radio fluxes in the cores of radio sources at the frequency of $8$ GHz, for the H21+ source list, we run an additional test with an extra factor $F^{X}_{b}$ in Eq. (\ref{eq:weights}), where $F^{X}_{b}$ is the relative radio flux of sources in the X radio band (between 8 GHz and 12 GHz). From the RFC, we use the column with the ``median flux density at X-band at baseline projection lengths shorter than 1000 km'' \citep{Petrov2025ApJS} and normalise it by the maximum value among the sources in the H21+ source list. This test does not result in a significant correlation with a p-value curve similar to the one in Fig.~\ref{fig:p_value_plot}; see Fig.~\ref{fig:p_value_plot_homan_radio} and Table~\ref{tab:top_ten_Homan_radio} in the Appendix.
    
\section{Discussion and conclusion}\label{sec:discussion}

    The two statistical tests weighing blazar-associated neutrino alerts accounting for the time lag of the neutrino arrival due to slowness of the proton energy losses in $p\gamma$ interactions and/or proton acceleration resulted in no evidence of a significant correlation between IceCat-1 alerts and the AGNs from H21+ or R24+ source lists. The low statistical significance of these associations $(\lesssim 2 \sigma)$, given the small size, is in line with predictions of \citet{Liodakis:2022ccz} and implies \citep{IceCube:2023htm,Kouch:2025dgz,Kouch:2025ipk} that only a small fraction of AGNs can be associated with \textit{IceCube} alerts.

    A set of non-mutually-exclusive reasons can explain the null results of our search:
    \begin{itemize}
        \item Blazars are inefficient sources of $\sim 100$~TeV neutrinos. Indeed, in our recent work \citep{Podlesnyi:2025aqb}, we estimated from an extrapolation of a leptohadronic model for the brightest blazar flare of 3C~454.3 to the whole population of \textit{Fermi}-LAT FSRQs that, on long-term average, they produce $\sim\left(\rho_{p/e} /130\right) \times 0.5 $\% of IceCube neutrinos at energies greater than $100$~TeV, where $\rho_{p/e}$ is the ratio of the proton and electron energy densities, which is a proxy of the baryon loading factor (the ratio of the proton and $\gamma$-ray luminosities). Taking this rough estimate at face value, given $N_{\nu} = 348$ IceCat-1 alerts with their median signalness of $0.41$, we obtain $0.71$ expected neutrinos from \textit{Fermi}-LAT FSRQs, which naively implies $\mathcal{O}(1)$ AGN-neutrino associations, assuming that other types of \textit{Fermi}-LAT AGNs contribute to the neutrino flux similarly.
        \item Blazars are efficient neutrino sources at energies $\gtrsim 10$~PeV. If blazars are efficient accelerators of protons\footnote{By this we mean the case when protons are accelerated close to the Hillas limit \citep{Hillas:1984ijl}, when the ratio of the proton acceleration timescale to the gyroperiod at the maximum energy $\xi_{\mathrm{acc}} \lesssim 10^{3}$.}, then the peak of the neutrino spectral-energy distributions will be around $E_{\nu} \gtrsim 100$~PeV because the protons of similar energies in the jet frame reach the $p\gamma$ reaction threshold and begin efficiently interacting with the abundant infrared photons (see, e.g., Fig.~12 by \citealp{Podlesnyi:2025aqb}). \citet{Rodrigues:2025cpm} also shows that this allows one to decrease the proton power below the Eddington limit, alleviating the issue with super-Eddington proton luminosities often required to explain $\sim 100$~TeV neutrinos.
        \item Our initial assumption of the universality of the neutrino time delay $t^{\prime}_{\mathrm{delay}}$ in the blazar jet frame may not be valid. Indeed, the conditions in different sources (e.g., acceleration mechanisms and efficiencies, magnetic and photon fields etc.) can vary too much from source to source and the universal value of $t^{\prime}_{\mathrm{delay}}$ may not exist.
        \item Even if $t^{\prime}_{\mathrm{delay}}$ is universal for all neutrino sources, its transformation from the jet frame to the observer frame as $t^{\prime}_{\mathrm{delay}} (1 + z) / D$ is significantly affected by the value of $D$. Indeed, there are $174$ AGNs which are present both in the H21+ and R24+ source lists, and the mean ratio of values from H21+ and R24+ is $\langle D^{\mathrm{H}21+} /  D^{\mathrm{R}24+} \rangle = 1.8$ and the standard deviation of $\sigma (D^{\mathrm{H}21+} /  D^{\mathrm{R}24+}) = 2.1$ indicating a rather large discrepancy between the values obtained with the two different methods (clearly visible also in Figs.~\ref{fig:doppler} and~\ref{fig:scatter}). The differences in estimates of Doppler factors in blazar jets with different methods are well known \citep{Liodakis:2015oea}, and even within the same method, at different observational epochs, the same source may have various values of $D$ \citep{Jorstad:2013vga}. Our adoption of the Gaussian weight accounting for the delay was meant to allow some room for uncertainty in $D$, but if it is as large as $\sim 100$\%, this can hinder attempts to reveal a significant correlation between major $\gamma$-ray flares and arrivals of \textit{IceCube} neutrinos.
        \item Proton acceleration and/or energy dissipation are decoupled from those of electrons. As \citet{Capel:2022cnm} pointed out, in many leptohadronic models, the hadronic contribution is not needed to describe the observational electromagnetic data well \citep{Keivani:2018rnh,MAGIC:2018sak,Murase:2018iyl,Sahakyan:2018voh,Gao:2018mnu,Oikonomou:2019djc,Cerruti:2018tmc,Reimer:2018vvw,Rodrigues:2018tku,Petropoulou:2019zqp,Oikonomou:2021akf}. This implies that potential neutrino emission even if some hadrons are present in the jet might be not causally connected to the leptonic emission because, e.g., protons are accelerated in a different region of the jet and/or most efficiently dissipate in zones different from those of electrons. This, in turn, leads to (i) neutrino flux not necessarily related to the electromagnetic flux of the source; (ii) no characteristic time delay $t^{\prime}_{\mathrm{delay}}$ between the neutrino production and electromagnetic flares.
    \end{itemize}
    
    \citet{Plavin:2020emb,Hovatta:2020lor,Kouch:2024xtd} found an indication of the association of \textit{IceCube} high-energy neutrinos with radio flares (see, however, recent work by \citet{Abbasi:2025bpg}, where such a correlation was not found), which are observed, on average, to lag behind $\gamma$-ray flares by several months in the Earth frame \citep{Kramarenko:2021rkf} as expected in some models \citep{Boula:2021pud,Tramacere:2021lug,Zacharias:2022spe}. Future studies utilising both $\gamma$-ray and radio light curves may increase the sensitivity of searches for neutrino-blazar associations. The releases of the new CAZ catalogue with $858$ Doppler factors \citep{Kouch:2025uwz} and the updated IceCat-2 catalogue of \textit{IceCube} alerts \citep{IceCube:2025uzh} will significantly enlarge both the sizes of the ensembles of sources and neutrino alerts, allowing to improve the sensitivity of the search for potentially delayed neutrinos.
    
    Other projected or constructed neutrino telescopes, such as the Pacific Ocean Neutrino Experiment \citep{P-ONE:2020ljt}, the Baikal Gigaton Volume Detector \citep{Avrorin:2011zzc}, and the Cubic Kilometre Neutrino Telescope \citep{KM3Net:2016zxf}, Tropical Deep-sea Neutrino Telescope \citep{TRIDENT:2022hql}, \textit{IceCube}-Gen2 \citep{IceCube-Gen2:2020qha,IceCube-Gen2:2023vtj}, and the Radio Neutrino Observatory in Greenland \citep{RNO-G:2020rmc} will help to provide the definitive answer to the question about the total contribution of the AGNs to the astrophysical neutrino flux and a role of the possible neutrino delay with respect to the electromagnetic emission.

    In conclusion, the two cross-matching tests between IceCat-1 neutrino alerts and AGNs from the MOJAVE-based source sample H21+ \citep{Homan:2021ijm} and CGRaBS-based source sample R24+ \citep{Rodrigues:2023vbv} with scanning over the jet-frame time delay $t^{\prime}_{\mathrm{delay}}$ between the neutrino arrival time in \textit{IceCube} and prior major $\gamma$-ray flares resulted in no significant evidence of the correlation.

\section*{Acknowledgments}

We thank the anonymous Referee for their comments, which helped us to improve the quality of the paper. We thank Nikita Vasiliev for multiple discussions during his earlier unpublished work on the student project on searching for spatio-temporal correlations between AGN $\gamma$-ray emission and \textit{IceCube} neutrinos. We thank Sergey Troitsky, Yuri Kovalev, and Timur Dzhatdoev for helpful discussions.

This research has made use of the \textit{Fermi}-LAT \citep{Fermi-LAT:2009ihh,Fermi-LAT:2019pir,Fermi-LAT:2019yla,Fermi-LAT:2023iml} data provided by NASA Goddard Space Flight Center, and of the NASA/IPAC Extragalactic Database, which is funded by the National Aeronautics and Space Administration and operated by the California Institute of Technology. This research has made use of data from the MOJAVE database, which is maintained by the MOJAVE team \citep{Lister2018ApJS}, and the data from the All-Sky Survey of Gamma-Ray Blazar Candidates CGRaBS \citep{Healey:2007gb}. This work has made use of the following software packages: \texttt{Python} \citep{python}, \texttt{AstroPy} \citep{astropy:2013, astropy:2018, astropy:2022}, \texttt{Jupyter} \citep{2007CSE.....9c..21P, kluyver2016jupyter}, \texttt{matplotlib} \citep{Hunter:2007}, \texttt{NumPy} \citep{numpy}, \texttt{SciPy} \citep{2020SciPy-NMeth, scipy_4718897}, \texttt{astroquery} \citep{2019AJ....157...98G, astroquery_10799414}, \texttt{pandas} \citep{pandasA,pandasB}, \texttt{snakemake} \citep{Snakemake}, \texttt{pyLCR} \citep{Fermi-LAT:2023iml}\footnote{\url{https://github.com/dankocevski/pyLCR}}, \texttt{matplotlib-venn}\footnote{\url{https://github.com/konstantint/matplotlib-venn}}, and \texttt{tqdm} \citep{tqdm_8233425}. This research has made use of NASA's Astrophysics Data System, INSPIRE-HEP, alphaXiv, and Pathfinder \citep{pathfinder}. Software citation information aggregated using the Software Citation Station \citep{software-citation-station-zenodo,software-citation-station-paper}\footnote{\url{https://www.tomwagg.com/software-citation-station/}}.

\section*{Data availability}
The compiled source lists and all the codes reproducing the results presented in the paper are available on \textit{Zenodo}: \cite{PodlesnyiOikonomou2026Zenodo}.

\bibliographystyle{aasjournal}

\bibliography{main_OJA}

@article{IceCube:2023htm,
    author = "Abbasi, R. and others",
    collaboration = "IceCube",
    title = "{Search for Correlations of High-energy Neutrinos Detected in IceCube with Radio-bright AGN and Gamma-Ray Emission from Blazars}",
    eprint = "2304.12675",
    archivePrefix = "arXiv",
    primaryClass = "astro-ph.HE",
    doi = "10.3847/1538-4357/acdfcb",
    journal = "Astrophys. J.",
    volume = "954",
    number = "1",
    pages = "75",
    year = "2023"
}

@article{IceCube:2023agq,
    author = "Abbasi, R. and others",
    collaboration = "IceCube",
    title = "{IceCat-1: The IceCube Event Catalog of Alert Tracks}",
    eprint = "2304.01174",
    archivePrefix = "arXiv",
    primaryClass = "astro-ph.HE",
    doi = "10.3847/1538-4365/acfa95",
    journal = "Astrophys. J. Suppl.",
    volume = "269",
    number = "1",
    pages = "25",
    year = "2023"
}

@article{Fermi-LAT:2019yla,
    author = "Abdollahi, S. and others",
    collaboration = "Fermi-LAT",
    title = "{$Fermi$ Large Area Telescope Fourth Source Catalog}",
    eprint = "1902.10045",
    archivePrefix = "arXiv",
    primaryClass = "astro-ph.HE",
    doi = "10.3847/1538-4365/ab6bcb",
    journal = "Astrophys. J. Suppl.",
    volume = "247",
    number = "1",
    pages = "33",
    year = "2020"
}

@article{Fermi-LAT:2023iml,
    author = "Abdollahi, S. and others",
    collaboration = "Fermi-LAT",
    title = "{The Fermi-LAT Lightcurve Repository}",
    eprint = "2301.01607",
    archivePrefix = "arXiv",
    primaryClass = "astro-ph.HE",
    doi = "10.3847/1538-4365/acbb6a",
    journal = "Astrophys. J. Suppl.",
    volume = "265",
    number = "2",
    pages = "31",
    year = "2023"
}

@article{Podlesnyi:2025aqb,
    author = {{Podlesnyi}, Egor and {Oikonomou}, Foteini},
        title = "{Insights from leptohadronic modelling of the brightest blazar flare}",
      journal = {Mon. Not. Roy. Astron. Soc.},
         year = 2025,
        month = oct,
          doi = {10.1093/mnras/staf1779},
archivePrefix = {arXiv},
       eprint = {2502.12111},
}

@article{Kouch:2025dgz,
    author = "Kouch, Pouya M. and Lindfors, Elina and Hovatta, Talvikki and Liodakis, Ioannis and Koljonen, Karri I. I. and Nilsson, Kari and Jormanainen, Jenni and Fallah Ramazani, Vandad and Graham, Matthew J.",
    title = "{Optimizing the hunt for extraterrestrial high-energy neutrino counterparts}",
    eprint = "2502.17567",
    archivePrefix = "arXiv",
    primaryClass = "astro-ph.HE",
    doi = "10.1051/0004-6361/202453277",
    journal = "Astron. Astrophys.",
    volume = "696",
    pages = "A73",
    year = "2025"
}

@article{Plavin:2022oyy,
    author = "Plavin, A. V. and Kovalev, Y. Y. and Kovalev, Yu A. and Troitsky, S. V.",
    title = "{Growing evidence for high-energy neutrinos originating in radio blazars}",
    eprint = "2211.09631",
    archivePrefix = "arXiv",
    primaryClass = "astro-ph.HE",
    reportNumber = "INR-TH-2022-024",
    doi = "10.1093/mnras/stad1467",
    journal = "Mon. Not. Roy. Astron. Soc.",
    volume = "523",
    number = "2",
    pages = "1799--1808",
    year = "2023"
}

@ARTICLE{Petrov2025ApJS,
       author = {{Petrov}, L.~Y. and {Kovalev}, Y.~Y.},
        title = "{The Radio Fundamental Catalog. I. Astrometry}",
      journal = {Astrophys. J. Suppl.},
         year = 2025,
        month = feb,
       volume = {276},
       number = {2},
          eid = {38},
        pages = {38},
          doi = {10.3847/1538-4365/ad8c36},
archivePrefix = {arXiv},
       eprint = {2410.11794},
 primaryClass = {astro-ph.IM},
       adsurl = {https://ui.adsabs.harvard.edu/abs/2025ApJS..276...38P}
}

@article{Plavin:2020emb,
    author = "Plavin, Alexander and Kovalev, Yuri Y. and Kovalev, Yuri A. and Troitsky, Sergey",
    title = "{Observational Evidence for the Origin of High-energy Neutrinos in Parsec-scale Nuclei of Radio-bright Active Galaxies}",
    eprint = "2001.00930",
    archivePrefix = "arXiv",
    primaryClass = "astro-ph.HE",
    reportNumber = "INR-TH-2020-002",
    doi = "10.3847/1538-4357/ab86bd",
    journal = "Astrophys. J.",
    volume = "894",
    number = "2",
    pages = "101",
    year = "2020"
}

@article{Plavin:2020mkf,
    author = "Plavin, A. V. and Kovalev, Y. Y. and Kovalev, Yu. A. and Troitsky, S. V.",
    title = "{Directional Association of TeV to PeV Astrophysical Neutrinos with Radio Blazars}",
    eprint = "2009.08914",
    archivePrefix = "arXiv",
    primaryClass = "astro-ph.HE",
    reportNumber = "INR-TH-2020-039",
    doi = "10.3847/1538-4357/abceb8",
    journal = "Astrophys. J.",
    volume = "908",
    number = "2",
    pages = "157",
    year = "2021"
}

@article{Homan:2021ijm,
    author = "Homan, D. C. and Cohen, M. H. and Hovatta, T. and Kellermann, K. I. and Kovalev, Y. Y. and Lister, M. L. and Popkov, A. V. and Pushkarev, A. B. and Ros, E. and Savolainen, T.",
    title = "{MOJAVE. XIX. Brightness Temperatures and Intrinsic Properties of Blazar Jets}",
    eprint = "2109.04977",
    archivePrefix = "arXiv",
    primaryClass = "astro-ph.HE",
    doi = "10.3847/1538-4357/ac27af",
    journal = "Astrophys. J.",
    volume = "923",
    number = "1",
    pages = "67",
    year = "2021"
}

@article{Fermi-LAT:2019pir,
    author = "Ajello, M. and others",
    collaboration = "Fermi-LAT",
    title = "{The Fourth Catalog of Active Galactic Nuclei Detected by the Fermi Large Area Telescope}",
    eprint = "1905.10771",
    archivePrefix = "arXiv",
    primaryClass = "astro-ph.HE",
    doi = "10.3847/1538-4357/ab791e",
    journal = "Astrophys. J.",
    volume = "892",
    pages = "105",
    year = "2020"
}

@article{Healey:2007gb,
    author = "Healey, Stephen Edward and Romani, Roger W. and Cotter, Garret and Michelson, Peter F. and Schlafly, Edward F. and Readhead, Anthony C. S. and Giommi, Paolo and Chaty, Sylvain and Grenier, Isabelle A. and Weintraub, Lawrence C.",
    title = "{CGRaBS: An All-Sky Survey of Gamma-Ray Blazar Candidates}",
    eprint = "0709.1735",
    archivePrefix = "arXiv",
    primaryClass = "astro-ph",
    doi = "10.1086/523302",
    journal = "Astrophys. J. Suppl.",
    volume = "175",
    pages = "97",
    year = "2008"
}

@article{Kouch:2024xtd,
    author = "Kouch, Pouya M. and others",
    title = "{Association of the IceCube neutrinos with blazars in the CGRaBS sample}",
    eprint = "2407.07153",
    archivePrefix = "arXiv",
    primaryClass = "astro-ph.HE",
    doi = "10.1051/0004-6361/202347624",
    journal = "Astron. Astrophys.",
    volume = "690",
    pages = "A111",
    year = "2024"
}

@book{Davison_Hinkley_1997,
place={Cambridge},
series={Cambridge Series in Statistical and Probabilistic Mathematics},
title={Bootstrap Methods and their Application},
publisher={Cambridge University Press},
author={Davison, A. C. and Hinkley, D. V.},
year={1997},
collection={Cambridge Series in Statistical and Probabilistic Mathematics}
}

@article{Kreter:2020kpm,
    author = {Kreter, M. and Kadler, M. and Krau\ss{}, F. and Mannheim, K. and Buson, S. and Ojha, R. and Wilms, J. and B\"ottcher, M.},
    title = "{On the Detection Potential of Blazar Flares for Current Neutrino Telescopes}",
    eprint = "2009.00125",
    archivePrefix = "arXiv",
    primaryClass = "astro-ph.HE",
    doi = "10.3847/1538-4357/abb5b1",
    journal = "Astrophys. J.",
    volume = "902",
    number = "2",
    pages = "133",
    year = "2020"
}

@article{Oikonomou:2019djc,
    author = "Oikonomou, Foteini and Murase, Kohta and Padovani, Paolo and Resconi, Elisa and M\'esz\'aros, Peter",
    title = "{High energy neutrino flux from individual blazar flares}",
    eprint = "1906.05302",
    archivePrefix = "arXiv",
    primaryClass = "astro-ph.HE",
    doi = "10.1093/mnras/stz2246",
    journal = "Mon. Not. Roy. Astron. Soc.",
    volume = "489",
    number = "3",
    pages = "4347--4366",
    year = "2019"
}

@article{Fermi-LAT:2011iez,
    author = {{Fermi-LAT Collaboration}},
    title = {Fermi Gamma-ray Space Telescope Observations of the Gamma-ray Outburst from 3C 454.3 in November 2010},
    eprint = {1102.0277},
    archivePrefix = {arXiv},
    primaryClass = {astro-ph.HE},
    doi = {10.1088/2041-8205/733/2/L26},
    journal = {Astrophys. J. Lett.},
    volume = {733},
    pages = {L26},
    year = {2011}
}

@article{IceCube:2018dnn,
    author = "Aartsen, M. G. and others",
    collaboration = "IceCube, Fermi-LAT, MAGIC, AGILE, ASAS-SN, HAWC, H.E.S.S., INTEGRAL, Kanata, Kiso, Kapteyn, Liverpool Telescope, Subaru, Swift NuSTAR, VERITAS, VLA/17B-403",
    title = "{Multimessenger observations of a flaring blazar coincident with high-energy neutrino IceCube-170922A}",
    eprint = "1807.08816",
    archivePrefix = "arXiv",
    primaryClass = "astro-ph.HE",
    doi = "10.1126/science.aat1378",
    journal = "Science",
    volume = "361",
    number = "6398",
    pages = "eaat1378",
    year = "2018"
}

@article{Keivani:2018rnh,
    author = "Keivani, A. and others",
    title = "{A Multimessenger Picture of the Flaring Blazar TXS 0506+056: implications for High-Energy Neutrino Emission and Cosmic Ray Acceleration}",
    eprint = "1807.04537",
    archivePrefix = "arXiv",
    primaryClass = "astro-ph.HE",
    doi = "10.3847/1538-4357/aad59a",
    journal = "Astrophys. J.",
    volume = "864",
    number = "1",
    pages = "84",
    year = "2018"
}

@article{MAGIC:2018sak,
    author = "Ansoldi, S. and others",
    collaboration = "MAGIC",
    title = "{The blazar TXS 0506+056 associated with a high-energy neutrino: insights into extragalactic jets and cosmic ray acceleration}",
    eprint = "1807.04300",
    archivePrefix = "arXiv",
    primaryClass = "astro-ph.HE",
    doi = "10.3847/2041-8213/aad083",
    journal = "Astrophys. J. Lett.",
    volume = "863",
    pages = "L10",
    year = "2018"
}

@article{Murase:2018iyl,
    author = "Murase, Kohta and Oikonomou, Foteini and Petropoulou, Maria",
    title = "{Blazar Flares as an Origin of High-Energy Cosmic Neutrinos?}",
    eprint = "1807.04748",
    archivePrefix = "arXiv",
    primaryClass = "astro-ph.HE",
    doi = "10.3847/1538-4357/aada00",
    journal = "Astrophys. J.",
    volume = "865",
    number = "2",
    pages = "124",
    year = "2018"
}

@article{Sahakyan:2018voh,
    author = "Sahakyan, N.",
    title = "{Lepto-hadronic $\gamma$-ray and neutrino emission from the jet of TXS 0506+056}",
    eprint = "1808.05651",
    archivePrefix = "arXiv",
    primaryClass = "astro-ph.HE",
    doi = "10.3847/1538-4357/aadade",
    journal = "Astrophys. J.",
    volume = "866",
    number = "2",
    pages = "109",
    year = "2018"
}

@article{Gao:2018mnu,
    author = "Gao, Shan and Fedynitch, Anatoli and Winter, Walter and Pohl, Martin",
    title = "{Modelling the coincident observation of a high-energy neutrino and a bright blazar flare}",
    eprint = "1807.04275",
    archivePrefix = "arXiv",
    primaryClass = "astro-ph.HE",
    doi = "10.1038/s41550-018-0610-1",
    journal = "Nature Astron.",
    volume = "3",
    number = "1",
    pages = "88--92",
    year = "2019"
}

@article{Cerruti:2018tmc,
    author = "Cerruti, M. and Zech, A. and Boisson, C. and Emery, G. and Inoue, S. and Lenain, J. -P.",
    title = "{Leptohadronic single-zone models for the electromagnetic and neutrino emission of TXS 0506+056}",
    eprint = "1807.04335",
    archivePrefix = "arXiv",
    primaryClass = "astro-ph.HE",
    doi = "10.1093/mnrasl/sly210",
    journal = "Mon. Not. Roy. Astron. Soc.",
    volume = "483",
    number = "1",
    pages = "L12--L16",
    year = "2019",
    note = "[Erratum: Mon.Not.Roy.Astron.Soc. 502, L21--L22 (2021)]"
}

@article{Reimer:2018vvw,
    author = "Reimer, Anita and Boettcher, Markus and Buson, Sara",
    title = "{Cascading Constraints from Neutrino-emitting Blazars: The Case of TXS 0506+056}",
    eprint = "1812.05654",
    archivePrefix = "arXiv",
    primaryClass = "astro-ph.HE",
    doi = "10.3847/1538-4357/ab2bff",
    journal = "Astrophys. J.",
    volume = "881",
    number = "1",
    pages = "46",
    year = "2019",
    note = "[Erratum: Astrophys.J. 899, 168 (2020)]"
}

@article{Rodrigues:2018tku,
    author = "Rodrigues, Xavier and Gao, Shan and Fedynitch, Anatoli and Palladino, Andrea and Winter, Walter",
    title = "{Leptohadronic Blazar Models Applied to the 2014\textendash{}2015 Flare of TXS 0506+056}",
    eprint = "1812.05939",
    archivePrefix = "arXiv",
    primaryClass = "astro-ph.HE",
    reportNumber = "DESY-19-019",
    doi = "10.3847/2041-8213/ab1267",
    journal = "Astrophys. J. Lett.",
    volume = "874",
    number = "2",
    pages = "L29",
    year = "2019"
}

@article{Petropoulou:2019zqp,
    author = "Petropoulou, Maria and others",
    title = "{Multi-Epoch Modeling of TXS 0506+056 and Implications for Long-Term High-Energy Neutrino Emission}",
    eprint = "1911.04010",
    archivePrefix = "arXiv",
    primaryClass = "astro-ph.HE",
    doi = "10.3847/1538-4357/ab76d0",
    journal = "Astrophys. J.",
    volume = "891",
    pages = "115",
    year = "2020"
}

@article{Plavin:2025pjt,
    author = "Plavin, Alexander V. and Kovalev, Yuri Y. and Troitsky, Sergey V.",
    title = "{Extreme Jet Beaming Observed in Neutrino-associated Blazars}",
    eprint = "2503.08667",
    archivePrefix = "arXiv",
    primaryClass = "astro-ph.HE",
    reportNumber = "INR-TH-2025-002",
    doi = "10.3847/1538-4357/adf54f",
    journal = "Astrophys. J.",
    volume = "991",
    number = "1",
    pages = "33",
    year = "2025"
}

@article{Oikonomou:2021akf,
    author = "Oikonomou, Foteini and Petropoulou, Maria and Murase, Kohta and Tohuvavohu, Aaron and Vasilopoulos, Georgios and Buson, Sara and Santander, Marcos",
    title = "{Multi-messenger emission from the parsec-scale jet of the flat-spectrum radio quasar coincident with high-energy neutrino IceCube-190730A}",
    eprint = "2107.11437",
    archivePrefix = "arXiv",
    primaryClass = "astro-ph.HE",
    doi = "10.1088/1475-7516/2021/10/082",
    journal = "JCAP",
    volume = "10",
    pages = "082",
    year = "2021"
}

@article{Kovalev:2025kxf,
    author = "Kovalev, Y. Y. and Pushkarev, A. B. and Gomez, J. L. and Homan, D. C. and Lister, M. L. and Livingston, J. D. and Pashchenko, I. N. and Plavin, A. V. and Savolainen, T. and Troitsky, S. V.",
    title = "{Looking into the jet cone of the neutrino-associated very high-energy blazar PKS 1424+240}",
    eprint = "2504.09287",
    archivePrefix = "arXiv",
    primaryClass = "astro-ph.HE",
    reportNumber = "INR-TH-2025-004",
    doi = "10.1051/0004-6361/202555400",
    journal = "Astron. Astrophys.",
    volume = "700",
    pages = "L12",
    year = "2025"
}

@article{Liodakis:2022ccz,
    author = "Liodakis, I. and Hovatta, T. and Pavlidou, V. and Readhead, A. C. S. and Blandford, R. D. and Kiehlmann, S. and Lindfors, E. and Max-Moerbeck, W. and Pearson, T. J. and Petropoulou, M.",
    title = "{The hunt for extraterrestrial high-energy neutrino counterparts}",
    eprint = "2208.07381",
    archivePrefix = "arXiv",
    primaryClass = "astro-ph.HE",
    doi = "10.1051/0004-6361/202244551",
    journal = "Astron. Astrophys.",
    volume = "666",
    pages = "A36",
    year = "2022"
}

@article{Hovatta:2020lor,
    author = "Hovatta, T. and others",
    title = {{Association of IceCube neutrinos with radio sources observed at Owens Valley and Mets\"ahovi Radio Observatories}},
    eprint = "2009.10523",
    archivePrefix = "arXiv",
    primaryClass = "astro-ph.HE",
    doi = "10.1051/0004-6361/202039481",
    journal = "Astron. Astrophys.",
    volume = "650",
    pages = "A83",
    year = "2021"
}

@article{Kovalev:2005yd,
    author = "Kovalev, Yuri Y. and others",
    title = "{Sub-milliarcsecond imaging of quasars and active galactic nuclei. 4. Fine scale structure}",
    eprint = "astro-ph/0505536",
    archivePrefix = "arXiv",
    doi = "10.1086/497430",
    journal = "Astron. J.",
    volume = "130",
    pages = "2473--2505",
    year = "2005"
}

@article{astropy:2013,
Adsurl = {http://adsabs.harvard.edu/abs/2013A%26A...558A..33A},
Archiveprefix = {arXiv},
Author = {{Astropy Collaboration}},
Doi = {10.1051/0004-6361/201322068},
Eid = {A33},
Eprint = {1307.6212},
Journal = {Astron. J.ap},
Month = oct,
Pages = {A33},
Primaryclass = {astro-ph.IM},
Title = {{Astropy: A community Python package for astronomy}},
Volume = 558,
Year = 2013}

@ARTICLE{astropy:2018,
       author = {{Astropy Collaboration} and {Astropy Contributors}},
        title = "{The Astropy Project: Building an Open-science Project and Status of the v2.0 Core Package}",
      journal = {Astron. J.},
         year = 2018,
        month = sep,
       volume = {156},
       number = {3},
          eid = {123},
        pages = {123},
          doi = {10.3847/1538-3881/aabc4f},
archivePrefix = {arXiv},
       eprint = {1801.02634},
 primaryClass = {astro-ph.IM},
       adsurl = {https://ui.adsabs.harvard.edu/abs/2018AJ....156..123A}
}

@ARTICLE{astropy:2022,
       author = {{Astropy Collaboration}},
        title = "{The Astropy Project: Sustaining and Growing a Community-oriented Open-source Project and the Latest Major Release (v5.0) of the Core Package}",
      journal = {Astrophys. J.},
         year = 2022,
        month = aug,
       volume = {935},
       number = {2},
          eid = {167},
        pages = {167},
          doi = {10.3847/1538-4357/ac7c74},
archivePrefix = {arXiv},
       eprint = {2206.14220},
 primaryClass = {astro-ph.IM},
       adsurl = {https://ui.adsabs.harvard.edu/abs/2022ApJ...935..167A}
}

@software{astroquery_10799414,
  author       = {{Adam Ginsburg et al.}},
  title        = {astropy/astroquery: v0.4.7},
  month        = mar,
  year         = 2024,
  publisher    = {Zenodo},
  version      = {v0.4.7},
  doi          = {10.5281/zenodo.10799414},
  url          = {https://doi.org/10.5281/zenodo.10799414}
}

@ARTICLE{2019AJ....157...98G,
   author = {{Astroquery collaboration} and {a subset of the astropy collaboration}
	},
    title = "{astroquery: An Astronomical Web-querying Package in Python}",
  journal = {Astron. J.},
archivePrefix = "arXiv",
   eprint = {1901.04520},
 primaryClass = "astro-ph.IM",
     year = 2019,
    month = mar,
   volume = 157,
      eid = {98},
    pages = {98},
      doi = {10.3847/1538-3881/aafc33},
   adsurl = {https://adsabs.harvard.edu/abs/2019AJ....157...98G}
}

@misc{NED,
  doi = {10.26132/NED1},
  url = {https://catcopy.ipac.caltech.edu/dois/doi.php?id=10.26132/NED1},
  author = {{NASA/IPAC Extragalactic Database}},
  title = {{NASA/IPAC Extragalactic Database (NED)}},
  publisher = {IPAC},
  year = {2019}
}

@ARTICLE{2007CSE.....9c..21P,
       author = {{Perez}, Fernando and {Granger}, Brian E.},
        title = "{IPython: A System for Interactive Scientific Computing}",
      journal = {Computing in Science and Engineering},
         year = "2007",
        month = "Jan",
       volume = {9},
       number = {3},
        pages = {21-29},
          doi = {10.1109/MCSE.2007.53},
       adsurl = {https://ui.adsabs.harvard.edu/abs/2007CSE.....9c..21P}
}

@inproceedings{kluyver2016jupyter,
  title={Jupyter Notebooks-a publishing format for reproducible computational workflows.},
  author={Kluyver, Thomas and Ragan-Kelley, Benjamin and P{\'e}rez, Fernando and Granger, Brian E and Bussonnier, Matthias and Frederic, Jonathan and Kelley, Kyle and Hamrick, Jessica B and Grout, Jason and Corlay, Sylvain and others},
  booktitle={ELPUB},
  pages={87--90},
  year={2016}
}

@Article{Hunter:2007,
  Author    = {Hunter, J. D.},
  Title     = {Matplotlib: A 2D graphics environment},
  Journal   = {Computing in Science \& Engineering},
  Volume    = {9},
  Number    = {3},
  Pages     = {90--95},
  publisher = {IEEE COMPUTER SOC},
  doi       = {10.1109/MCSE.2007.55},
  year      = 2007
}

@article{numpy,
    author = "Harris, Charles R. and others",
    title = "{Array programming with NumPy}",
    eprint = "2006.10256",
    archivePrefix = "arXiv",
    primaryClass = "cs.MS",
    doi = "10.1038/s41586-020-2649-2",
    journal = "Nature",
    volume = "585",
    number = "7825",
    pages = "357--362",
    year = "2020"
}

@book{python,
  author    = {Van Rossum, Guido and Drake, Fred L.},
  title     = {Python 3 Reference Manual},
  year      = {2009},
  isbn      = {1441412697},
  publisher = {CreateSpace},
  address   = {Scotts Valley, CA}
}

@software{scipy_4718897,
  author       = {{Pauli Virtanen et al.}},
  title        = {scipy/scipy: SciPy 1.6.3},
  month        = apr,
  year         = 2021,
  publisher    = {Zenodo},
  version      = {v1.6.3},
  doi          = {10.5281/zenodo.4718897},
  url          = {https://doi.org/10.5281/zenodo.4718897}
}

@ARTICLE{2020SciPy-NMeth,
    author = "Virtanen, Pauli and others",
    title = "{SciPy 1.0--Fundamental Algorithms for Scientific Computing in Python}",
    eprint = "1907.10121",
    archivePrefix = "arXiv",
    primaryClass = "cs.MS",
    doi = "10.1038/s41592-019-0686-2",
    journal = "Nature Meth.",
    volume = "17",
    pages = "261",
    year = "2020"
}

@software{tqdm_8233425,
  author       = {{Casper da Costa-Luis et al.}},
  title        = {{tqdm: A fast, Extensible Progress Bar for Python and CLI}},
  month        = aug,
  year         = 2023,
  publisher    = {Zenodo},
  version      = {v4.66.1},
  doi          = {10.5281/zenodo.8233425},
  url          = {https://doi.org/10.5281/zenodo.8233425}
}

@ARTICLE{software-citation-station-paper,
       author = {{Wagg}, Tom and {Broekgaarden}, Floor S.},
        title = "{Streamlining and standardizing software citations with The Software Citation Station}",
      journal = {arXiv e-prints},
         year = 2024,
        month = jun,
          eid = {arXiv:2406.04405},
        pages = {arXiv:2406.04405},
archivePrefix = {arXiv},
       eprint = {2406.04405},
 primaryClass = {astro-ph.IM},
       adsurl = {https://ui.adsabs.harvard.edu/abs/2024arXiv240604405W}
}

@software{software-citation-station-zenodo,
  author       = {Wagg, Tom and Broekgaarden, Floor},
  title        = {The Software Citation Station},
  month        = may,
  year         = 2024,
  publisher    = {Zenodo},
  doi          = {10.5281/zenodo.11292917},
  url          = {https://doi.org/10.5281/zenodo.11292917}
}

@article{Snakemake,
  doi = {10.12688/f1000research.29032.2},
  url = {https://doi.org/10.12688/f1000research.29032.2},
  year = {2021},
  month = apr,
  publisher = {F1000 Research Ltd},
  volume = {10},
  pages = {33},
  author = {Felix M\"{o}lder and Kim Philipp Jablonski and Brice Letcher and Michael B. Hall and Christopher H. Tomkins-Tinch and Vanessa Sochat and Jan Forster and Soohyun Lee and Sven O. Twardziok and Alexander Kanitz and Andreas Wilm and Manuel Holtgrewe and Sven Rahmann and Sven Nahnsen and Johannes K\"{o}ster},
  title = {Sustainable data analysis with Snakemake},
  journal = {F1000Research}
}

@software{pandasA,
    author       = {{pandas development team}},
    title        = {pandas-dev/pandas: Pandas},
    month        = feb,
    year         = 2020,
    publisher    = {Zenodo},
    doi          = {10.5281/zenodo.3509134},
    url          = {https://doi.org/10.5281/zenodo.3509134}
}

@InProceedings{pandasB,
  author    = { {W}es {M}c{K}inney },
  title     = { {D}ata {S}tructures for {S}tatistical {C}omputing in {P}ython },
  booktitle = { {P}roceedings of the 9th {P}ython in {S}cience {C}onference },
  pages     = { 56 - 61 },
  year      = { 2010 },
  editor    = { {S}t\'efan van der {W}alt and {J}arrod {M}illman },
  doi       = { 10.25080/Majora-92bf1922-00a }
}

@article{pathfinder,
      title={pathfinder: A Semantic Framework for Literature Review and Knowledge Discovery in Astronomy}, 
      author={{Kartheik G. Iyer et al.}},
      year={2024},
      eprint={2408.01556},
      archivePrefix={arXiv},
      primaryClass={astro-ph.IM},
      url={https://arxiv.org/abs/2408.01556}, 
}

@article{Nalewajko:2012yf,
    author = "Nalewajko, Krzysztof",
    title = "{The brightest gamma-ray flares of blazars}",
    eprint = "1211.0274",
    archivePrefix = "arXiv",
    primaryClass = "astro-ph.HE",
    doi = "10.1093/mnras/sts711",
    journal = "Mon. Not. Roy. Astron. Soc.",
    volume = "430",
    pages = "1324",
    year = "2013"
}

@article{Plavin:2023wsb,
    author = "Plavin, A. V. and Burenin, R. A. and Kovalev, Y. Y. and Lutovinov, A. A. and Starobinsky, A. A. and Troitsky, S. V. and Zakharov, E. I.",
    title = "{Hard X-ray emission from blazars associated with high-energy neutrinos}",
    eprint = "2306.00960",
    archivePrefix = "arXiv",
    primaryClass = "astro-ph.HE",
    reportNumber = "INR-TH-2023-007",
    doi = "10.1088/1475-7516/2024/05/133",
    journal = "JCAP",
    volume = "05",
    pages = "133",
    year = "2024"
}

@article{Kramarenko:2021rkf,
    author = "Kramarenko, I. G. and Pushkarev, A. B. and Kovalev, Y. Y. and Lister, M. L. and Hovatta, T. and Savolainen, T.",
    title = "{A decade of joint MOJAVE\textendash{}Fermi AGN monitoring: localization of the gamma-ray emission region}",
    eprint = "2106.08416",
    archivePrefix = "arXiv",
    primaryClass = "astro-ph.HE",
    doi = "10.1093/mnras/stab3358",
    journal = "Mon. Not. Roy. Astron. Soc.",
    volume = "510",
    number = "1",
    pages = "469--480",
    year = "2021"
}

@article{Boula:2021pud,
    author = "Boula, Stella S. and Mastichiadis, Apostolos",
    title = "{Expanding one-zone model for blazar emission}",
    eprint = "2110.05325",
    archivePrefix = "arXiv",
    primaryClass = "astro-ph.HE",
    doi = "10.1051/0004-6361/202142126",
    journal = "Astron. Astrophys.",
    volume = "657",
    pages = "A20",
    year = "2022"
}

@article{Tramacere:2021lug,
    author = "Tramacere, Andrea and Sliusar, Vitalii and Walter, Roland and Jurysek, Jakub and Balbo, Matteo",
    title = "{Radio-\ensuremath{\gamma}-ray response in blazars as a signature of adiabatic blob expansion}",
    eprint = "2112.03941",
    archivePrefix = "arXiv",
    primaryClass = "astro-ph.HE",
    doi = "10.1051/0004-6361/202142003",
    journal = "Astron. Astrophys.",
    volume = "658",
    pages = "A173",
    year = "2022"
}

@article{Zacharias:2022spe,
    author = "Zacharias, Michael",
    title = "{Exploring the evolution of the particle distribution and the cascade in a moving, expanding emission region in blazar jets}",
    eprint = "2211.12283",
    archivePrefix = "arXiv",
    primaryClass = "astro-ph.HE",
    doi = "10.1051/0004-6361/202244683",
    journal = "Astron. Astrophys.",
    volume = "669",
    pages = "A151",
    year = "2023"
}

@article{Lu:2025vmk,
    author = "Lu, Ming-Xuan and Liang, Yun-Feng and Wang, Xiang-Gao and Ouyang, Xue-Rui",
    title = "{Investigating the correlation between ZTF TDEs and IceCube high-energy neutrinos}",
    eprint = "2503.09426",
    archivePrefix = "arXiv",
    primaryClass = "astro-ph.HE",
    month = "3",
    year = "2025"
}

@article{Suray:2023lsa,
    author = "Suray, Alisa and Troitsky, Sergey",
    title = "{Neutrino flares of radio blazars observed from TeV to PeV}",
    eprint = "2306.16797",
    archivePrefix = "arXiv",
    primaryClass = "astro-ph.HE",
    reportNumber = "INR-TH-2023-010",
    doi = "10.1093/mnrasl/slad136",
    journal = "Mon. Not. Roy. Astron. Soc.",
    volume = "527",
    number = "1",
    pages = "L26--L31",
    year = "2023"
}

@article{Bellenghi:2023yza,
    author = "Bellenghi, Chiara and Padovani, Paolo and Resconi, Elisa and Giommi, Paolo",
    title = "{Correlating High-energy IceCube Neutrinos with 5BZCAT Blazars and RFC Sources}",
    eprint = "2309.03115",
    archivePrefix = "arXiv",
    primaryClass = "astro-ph.HE",
    doi = "10.3847/2041-8213/acf711",
    journal = "Astrophys. J. Lett.",
    volume = "955",
    number = "2",
    pages = "L32",
    year = "2023"
}

@article{Jorstad:2013vga,
    author = "Jorstad, Svetlana G. and others",
    title = "{A Tight Connection between Gamma-Ray Outbursts and Parsec-Scale Jet Activity in the Quasar 3C 454.3}",
    eprint = "1307.2522",
    archivePrefix = "arXiv",
    primaryClass = "astro-ph.HE",
    doi = "10.1088/0004-637X/773/2/147",
    journal = "Astrophys. J.",
    volume = "773",
    pages = "147",
    year = "2013"
}

@article{Buson:2022fyf,
    author = "Buson, Sara and others",
    title = "{Beginning a Journey Across the Universe: The Discovery of Extragalactic Neutrino Factories}",
    eprint = "2207.06314",
    archivePrefix = "arXiv",
    primaryClass = "astro-ph.HE",
    doi = "10.3847/2041-8213/ac7d5b",
    journal = "Astrophys. J. Lett.",
    volume = "933",
    number = "2",
    pages = "L43",
    year = "2022",
    note = "[Erratum: Astrophys.J.Lett. 934, L38 (2022), Erratum: Astrophys.J. 934, L38 (2022)]"
}

@ARTICLE{Buson:2022fyfErratum,
       author = {{Buson}, Sara and {Tramacere}, Andrea and {Pfeiffer}, Leonard and {Oswald}, Lenz and {de Menezes}, Raniere and {Azzollini}, Alessandra and {Ajello}, Marco},
        title = "{Erratum: ``Beginning a Journey Across the Universe: The Discovery of Extragalactic Neutrino Factories'' (2022, ApJL, 933, L43)}",
      journal = {Astrophys. J. Lett.},
         year = 2022,
        month = aug,
       volume = {934},
       number = {2},
          eid = {L38},
        pages = {L38},
          doi = {10.3847/2041-8213/ac83a2},
       adsurl = {https://ui.adsabs.harvard.edu/abs/2022ApJ...934L..38B}
}

@article{Robinson:2024hmz,
    author = "Robinson, Joshua and Boettcher, Markus",
    title = "{Neutrino Detection Rates from Lepto-hadronic Model Simulations of Bright Blazar Flares}",
    eprint = "2410.21881",
    archivePrefix = "arXiv",
    primaryClass = "astro-ph.HE",
    doi = "10.3847/1538-4357/ad8dce",
    journal = "Astrophys. J.",
    volume = "977",
    number = "1",
    pages = "42",
    year = "2024"
}

@article{IceCube:2013low,
    author = "Aartsen, M. G. and others",
    collaboration = "IceCube",
    title = "{Evidence for High-Energy Extraterrestrial Neutrinos at the IceCube Detector}",
    eprint = "1311.5238",
    archivePrefix = "arXiv",
    primaryClass = "astro-ph.HE",
    doi = "10.1126/science.1242856",
    journal = "Science",
    volume = "342",
    pages = "1242856",
    year = "2013"
}

@article{IceCube:2023ame,
    author = "Abbasi, R. and others",
    collaboration = "IceCube",
    title = "{Observation of high-energy neutrinos from the Galactic plane}",
    eprint = "2307.04427",
    archivePrefix = "arXiv",
    primaryClass = "astro-ph.HE",
    doi = "10.1126/science.adc9818",
    journal = "Science",
    volume = "380",
    number = "6652",
    pages = "adc9818",
    year = "2023"
}

@article{Kovalev:2022izi,
    author = "Kovalev, Y. Y. and Plavin, A. V. and Troitsky, S. V.",
    title = "{Galactic Contribution to the High-energy Neutrino Flux Found in Track-like IceCube Events}",
    eprint = "2208.08423",
    archivePrefix = "arXiv",
    primaryClass = "astro-ph.HE",
    reportNumber = "INR-TH-2022-018",
    doi = "10.3847/2041-8213/aca1ae",
    journal = "Astrophys. J. Lett.",
    volume = "940",
    number = "2",
    pages = "L41",
    year = "2022"
}

@article{IceCube:2022der,
    author = "Abbasi, R. and others",
    collaboration = "IceCube",
    title = "{Evidence for neutrino emission from the nearby active galaxy NGC 1068}",
    eprint = "2211.09972",
    archivePrefix = "arXiv",
    primaryClass = "astro-ph.HE",
    doi = "10.1126/science.abg3395",
    journal = "Science",
    volume = "378",
    number = "6619",
    pages = "538--543",
    year = "2022"
}

@article{Troitsky:2021nvu,
    author = "Troitsky, Sergey V.",
    title = "{Constraints on models of the origin of high-energy astrophysical neutrinos}",
    eprint = "2112.09611",
    archivePrefix = "arXiv",
    primaryClass = "astro-ph.HE",
    reportNumber = "INR-TH-2021-025",
    doi = "10.3367/UFNe.2021.09.039062",
    journal = "Usp. Fiz. Nauk",
    volume = "191",
    number = "12",
    pages = "1333--1360",
    year = "2021"
}

@article{Troitsky:2023nli,
    author = "Troitsky, Sergey",
    title = "{Origin of high-energy astrophysical neutrinos: new results and prospects}",
    eprint = "2311.00281",
    archivePrefix = "arXiv",
    primaryClass = "astro-ph.HE",
    reportNumber = "INR-TH-2023-018",
    doi = "10.3367/UFNr.2023.04.039581",
    journal = "Usp. Fiz. Nauk",
    volume = "194",
    number = "4",
    pages = "371--383",
    year = "2024"
}

@article{Baikal-GVD:2024kfx,
    author = "Allakhverdyan, V. A. and others",
    collaboration = "Baikal-GVD",
    title = "{Probing the Galactic Neutrino Flux at Neutrino Energies above 200 TeV with the Baikal Gigaton Volume Detector}",
    eprint = "2411.05608",
    archivePrefix = "arXiv",
    primaryClass = "astro-ph.HE",
    doi = "10.3847/1538-4357/adb630",
    journal = "Astrophys. J.",
    volume = "982",
    number = "2",
    pages = "73",
    year = "2025"
}

@article{Berezinsky81a,
    author = {Berezinsky, V. S. and Ginzburg, V. L.},
    title = {On high-energy neutrino radiation of quasars and active galactic nuclei},
    journal = {Monthly Notices of the Royal Astronomical Society},
    volume = {194},
    number = {1},
    pages = {3-14},
    year = {1981},
    month = {01},
    issn = {0035-8711},
    doi = {10.1093/mnras/194.1.3},
    url = {https://doi.org/10.1093/mnras/194.1.3},
    eprint = {https://academic.oup.com/mnras/article-pdf/194/1/3/3084834/mnras194-0003.pdf},
}

@article{ANTARES:2023lck,
    author = "Albert, A. and others",
    collaboration = "ANTARES, OVRO",
    title = "{Searches for Neutrinos in the Direction of Radio-bright Blazars with the ANTARES Telescope}",
    eprint = "2309.06874",
    archivePrefix = "arXiv",
    primaryClass = "astro-ph.HE",
    doi = "10.3847/1538-4357/ad1f5b",
    journal = "Astrophys. J.",
    volume = "964",
    number = "1",
    pages = "3",
    year = "2024"
}

@article{Dermer:2012rg,
    author = "Dermer, Charles D. and Murase, Kohta and Takami, Hajime",
    title = "{Variable Gamma-ray Emission Induced by Ultra-High Energy Neutral Beams: Application to 4C +21.35}",
    eprint = "1203.6544",
    archivePrefix = "arXiv",
    primaryClass = "astro-ph.HE",
    doi = "10.1088/0004-637X/755/2/147",
    journal = "Astrophys. J.",
    volume = "755",
    pages = "147",
    year = "2012"
}

@article{Giommi:2020hbx,
    author = "Giommi, P. and Glauch, T. and Padovani, P. and Resconi, E. and Turcati, A. and Chang, Y. L.",
    title = "{Dissecting the regions around IceCube high-energy neutrinos: growing evidence for the blazar connection}",
    eprint = "2001.09355",
    archivePrefix = "arXiv",
    primaryClass = "astro-ph.HE",
    doi = "10.1093/mnras/staa2082",
    journal = "Mon. Not. Roy. Astron. Soc.",
    volume = "497",
    number = "1",
    pages = "865--878",
    year = "2020"
}

@article{Rodrigues:2024fhu,
    author = "Rodrigues, Xavier and Karl, Martina and Padovani, Paolo and Giommi, Paolo and Paiano, Simona and Falomo, Renato and Petropoulou, Maria and Oikonomou, Foteini",
    title = "{The Spectra of IceCube Neutrino (SIN) candidate sources - V. Modeling and interpretation of multiwavelength and neutrino data}",
    eprint = "2406.06667",
    archivePrefix = "arXiv",
    primaryClass = "astro-ph.HE",
    doi = "10.1051/0004-6361/202450592",
    journal = "Astron. Astrophys.",
    volume = "689",
    pages = "A147",
    year = "2024"
}

@article{Rodrigues:2023vbv,
    author = "Rodrigues, Xavier and Paliya, Vaidehi S. and Garrappa, Simone and Omeliukh, Anastasiia and Franckowiak, Anna and Winter, Walter",
    title = "{Leptohadronic multi-messenger modeling of 324 gamma-ray blazars}",
    eprint = "2307.13024",
    archivePrefix = "arXiv",
    primaryClass = "astro-ph.HE",
    doi = "10.1051/0004-6361/202347540",
    journal = "Astron. Astrophys.",
    volume = "681",
    pages = "A119",
    year = "2024"
}

@data{IceCat-1v4,
author = {{IceCube Collaboration}},
publisher = {Harvard Dataverse},
title = {{ICECAT-1: IceCube Event Catalog of Alert Tracks}},
UNF = {UNF:6:vnU6oTF0vfq27WzKXCgM3A==},
year = {2023},
version = {V4},
doi = {10.7910/DVN/SCRUCD},
url = {https://doi.org/10.7910/DVN/SCRUCD}
}

@article{Neronov:2023aks,
    author = "Neronov, A. and Savchenko, D. and Semikoz, D. V.",
    title = "{Neutrino Signal from a Population of Seyfert Galaxies}",
    eprint = "2306.09018",
    archivePrefix = "arXiv",
    primaryClass = "astro-ph.HE",
    doi = "10.1103/PhysRevLett.132.101002",
    journal = "Phys. Rev. Lett.",
    volume = "132",
    number = "10",
    pages = "101002",
    year = "2024"
}

@article{Paliya:2017xaq,
    author = "Paliya, Vaidehi S. and Marcotulli, L. and Ajello, M. and Joshi, M. and Sahayanathan, S. and Rao, A. R. and Hartmann, D.",
    title = "{General Physical Properties of CGRaBS Blazars}",
    eprint = "1711.01292",
    archivePrefix = "arXiv",
    primaryClass = "astro-ph.HE",
    doi = "10.3847/1538-4357/aa98e1",
    journal = "Astrophys. J.",
    volume = "851",
    number = "1",
    pages = "33",
    year = "2017"
}

@article{Kuhlmann:2025ocn,
    author = "Kuhlmann, Julian and Capel, Francesca",
    title = "{Impact of multi-messenger spectral modelling on blazar-neutrino associations}",
    eprint = "2503.04632",
    archivePrefix = "arXiv",
    primaryClass = "astro-ph.HE",
    month = "3",
    year = "2025"
}

@article{Capel:2022cnm,
    author = "Capel, F. and Burgess, J. M. and Mortlock, D. J. and Padovani, P.",
    title = "{Assessing coincident neutrino detections using population models}",
    eprint = "2201.05633",
    archivePrefix = "arXiv",
    primaryClass = "astro-ph.HE",
    doi = "10.1051/0004-6361/202243116",
    journal = "Astron. Astrophys.",
    volume = "668",
    pages = "A190",
    year = "2022"
}

@article{Franckowiak:2020qrq,
    author = "Franckowiak, A. and others",
    title = "{Patterns in the Multiwavelength Behavior of Candidate Neutrino Blazars}",
    eprint = "2001.10232",
    archivePrefix = "arXiv",
    primaryClass = "astro-ph.HE",
    doi = "10.3847/1538-4357/ab8307",
    journal = "Astrophys. J.",
    volume = "893",
    number = "2",
    pages = "162",
    year = "2020"
}

@article{P-ONE:2020ljt,
    author = "Agostini, Matteo and others",
    collaboration = "P-ONE",
    title = "{The Pacific Ocean Neutrino Experiment}",
    eprint = "2005.09493",
    archivePrefix = "arXiv",
    primaryClass = "astro-ph.HE",
    doi = "10.1038/s41550-020-1182-4",
    journal = "Nature Astron.",
    volume = "4",
    number = "10",
    pages = "913--915",
    year = "2020"
}

@article{Avrorin:2011zzc,
    author = "Avrorin, A. and others",
    editor = "Forty, R. and Hallewell, G. and Hofmann, W. and Nappi, E. and Ratcliff, B.",
    title = "{The gigaton volume detector in Lake Baikal}",
    doi = "10.1016/j.nima.2010.09.137",
    journal = "Nucl. Instrum. Meth. A",
    volume = "639",
    pages = "30--32",
    year = "2011"
}

@article{KM3Net:2016zxf,
    author = "Adrian-Martinez, S. and others",
    collaboration = "KM3Net",
    title = "{Letter of intent for KM3NeT 2.0}",
    eprint = "1601.07459",
    archivePrefix = "arXiv",
    primaryClass = "astro-ph.IM",
    doi = "10.1088/0954-3899/43/8/084001",
    journal = "J. Phys. G",
    volume = "43",
    number = "8",
    pages = "084001",
    year = "2016"
}

@article{TRIDENT:2022hql,
    author = "Ye, Z. P. and others",
    collaboration = "TRIDENT",
    title = "{A multi-cubic-kilometre neutrino telescope in the western Pacific Ocean}",
    eprint = "2207.04519",
    archivePrefix = "arXiv",
    primaryClass = "astro-ph.HE",
    doi = "10.1038/s41550-023-02087-6",
    journal = "Nature Astron.",
    volume = "7",
    number = "12",
    pages = "1497--1505",
    year = "2023"
}

@article{IceCube-Gen2:2023vtj,
    author = "Ishihara, Aya and others",
    collaboration = "IceCube-Gen2",
    title = "{The next generation neutrino telescope: IceCube-Gen2}",
    eprint = "2308.09427",
    archivePrefix = "arXiv",
    primaryClass = "astro-ph.HE",
    reportNumber = "PoS-ICRC2023-994",
    doi = "10.22323/1.444.0994",
    journal = "PoS",
    volume = "ICRC2023",
    pages = "994",
    year = "2023"
}

@ARTICLE{Lister2018ApJS,
       author = {{Lister}, M.~L. and {Aller}, M.~F. and {Aller}, H.~D. and {Hodge}, M.~A. and {Homan}, D.~C. and {Kovalev}, Y.~Y. and {Pushkarev}, A.~B. and {Savolainen}, T.},
        title = "{MOJAVE. XV. VLBA 15 GHz Total Intensity and Polarization Maps of 437 Parsec-scale AGN Jets from 1996 to 2017}",
      journal = {Astrophys. J. Suppl.},
         year = 2018,
        month = jan,
       volume = {234},
       number = {1},
          eid = {12},
        pages = {12},
          doi = {10.3847/1538-4365/aa9c44},
archivePrefix = {arXiv},
       eprint = {1711.07802},
 primaryClass = {astro-ph.GA},
       adsurl = {https://ui.adsabs.harvard.edu/abs/2018ApJS..234...12L}
}

@article{Fermi-LAT:2009ihh,
    author = "Atwood, W. B. and others",
    collaboration = "Fermi-LAT",
    title = "{The Large Area Telescope on the Fermi Gamma-ray Space Telescope Mission}",
    eprint = "0902.1089",
    archivePrefix = "arXiv",
    primaryClass = "astro-ph.IM",
    reportNumber = "SLAC-PUB-13620",
    doi = "10.1088/0004-637X/697/2/1071",
    journal = "Astrophys. J.",
    volume = "697",
    pages = "1071--1102",
    year = "2009"
}

@article{Abbasi:2025bpg,
    author = "Abbasi, R. and others",
    title = "{A Search for Millimeter-Bright Blazars as Astrophysical Neutrino Sources}",
    eprint = "2507.03989",
    archivePrefix = "arXiv",
    primaryClass = "astro-ph.HE",
    month = "7",
    year = "2025"
}

@article{Kurahashi:2022utm,
    author = "Kurahashi, Naoko and Murase, Kohta and Santander, Marcos",
    title = "{High-Energy Extragalactic Neutrino Astrophysics}",
    eprint = "2203.11936",
    archivePrefix = "arXiv",
    primaryClass = "astro-ph.HE",
    doi = "10.1146/annurev-nucl-011122-061547",
    journal = "Ann. Rev. Nucl. Part. Sci.",
    volume = "72",
    pages = "365",
    year = "2022"
}

@article{KM3NeT2025,
  author    = "{KM3NeT Collaboration}",
  title     = "{Observation of an ultra-high-energy cosmic neutrino with KM3NeT}",
  journal   = "Nature",
  volume    = "638",
  pages     = "376--382",
  year      = "2025",
  doi       = "10.1038/s41586-024-08543-1",
  url       = "https://doi.org/10.1038/s41586-024-08543-1"
}

@article{Pittori:2018hmw,
    author = "Pittori, C. and others",
    title = "{The Bright $\gamma$-ray Flare of 3C 279 in June 2015: AGILE Detection and Multifrequency Follow-up Observations}",
    eprint = "1803.07529",
    archivePrefix = "arXiv",
    primaryClass = "astro-ph.HE",
    doi = "10.3847/1538-4357/aab1f9",
    journal = "Astrophys. J.",
    volume = "856",
    number = "2",
    pages = "99",
    year = "2018"
}

@article{Singh:2020upf,
    author = "Singh, K. K. and Meintjes, P. J. and Bisschoff, B. and Ramamonjisoa, F. A. and van Soelen, B.",
    title = "{Gamma-ray and Optical properties of the flat spectrum radio quasar 3C 279 flare in June 2015}",
    eprint = "2003.01999",
    archivePrefix = "arXiv",
    primaryClass = "astro-ph.HE",
    doi = "10.1016/j.jheap.2020.02.007",
    journal = "JHEAp",
    volume = "26",
    pages = "65--76",
    year = "2020"
}

@article{Paliya:2015tea,
    author = "Paliya, Vaidehi S.",
    title = "{Fermi-Large Area Telescope Observations of the Exceptional Gamma-ray Flare from 3C 279 in 2015 June}",
    eprint = "1507.03073",
    archivePrefix = "arXiv",
    primaryClass = "astro-ph.HE",
    doi = "10.1088/2041-8205/808/2/L48",
    journal = "Astrophys. J. Lett.",
    volume = "808",
    number = "2",
    pages = "L48",
    year = "2015"
}

@article{Liodakis:2015oea,
    author = "Liodakis, I. and Zezas, A. and Pavlidou, V. and Angelakis, E. and Hovatta, T.",
    title = "{Reconciling inverse-Compton Doppler factors with variability Doppler factors in blazar jets}",
    eprint = "1503.04780",
    archivePrefix = "arXiv",
    primaryClass = "astro-ph.HE",
    doi = "10.1051/0004-6361/201629902",
    journal = "Astron. Astrophys.",
    volume = "602",
    pages = "A104",
    year = "2017"
}

@article{IceCube-Gen2:2020qha,
    author = "Aartsen, M. G. and others",
    collaboration = "IceCube-Gen2",
    title = "{IceCube-Gen2: the window to the extreme Universe}",
    eprint = "2008.04323",
    archivePrefix = "arXiv",
    primaryClass = "astro-ph.HE",
    doi = "10.1088/1361-6471/abbd48",
    journal = "J. Phys. G",
    volume = "48",
    number = "6",
    pages = "060501",
    year = "2021"
}

@article{RNO-G:2020rmc,
    author = "Aguilar, J. A. and others",
    collaboration = "RNO-G",
    title = "{Design and Sensitivity of the Radio Neutrino Observatory in Greenland (RNO-G)}",
    eprint = "2010.12279",
    archivePrefix = "arXiv",
    primaryClass = "astro-ph.IM",
    doi = "10.1088/1748-0221/16/03/P03025",
    journal = "JINST",
    volume = "16",
    number = "03",
    pages = "P03025",
    year = "2021",
    note = "[Erratum: JINST 18, E03001 (2023)]"
}

@article{IceCube:2025uzh,
    author = "Abbasi, Rasha and others",
    collaboration = "IceCube",
    title = "{IceCat-2: Updated IceCube Event Catalog of Alert Tracks}",
    eprint = "2507.06176",
    archivePrefix = "arXiv",
    primaryClass = "astro-ph.HE",
    reportNumber = "PoS(ICRC2025)1224",
    doi = "10.22323/1.501.1224",
    journal = "PoS",
    volume = "ICRC2025",
    pages = "1224",
    year = "2025"
}

@article{Hillas:1984ijl,
    author = "Hillas, A. M.",
    title = "{The Origin of Ultrahigh-Energy Cosmic Rays}",
    doi = "10.1146/annurev.aa.22.090184.002233",
    journal = "Ann. Rev. Astron. Astrophys.",
    volume = "22",
    pages = "425--444",
    year = "1984"
}

@article{Ghisellini:2009fj,
    author = "Ghisellini, G. and Tavecchio, F. and Foschini, L. and Ghirlanda, G. and Maraschi, L. and Celotti, A.",
    title = "{General physical properties of bright Fermi blazars}",
    eprint = "0909.0932",
    archivePrefix = "arXiv",
    primaryClass = "astro-ph.CO",
    doi = "10.1111/j.1365-2966.2009.15898.x",
    journal = "Mon. Not. Roy. Astron. Soc.",
    volume = "402",
    pages = "497",
    year = "2010"
}

@article{Rodrigues:2025cpm,
    author = "Rodrigues, Xavier and Rieger, Frank and Bohdan, Artem and Padovani, Paolo",
    title = "{Hillas meets Eddington: The case for blazars as ultra-high-energy neutrino sources}",
    eprint = "2508.18345",
    archivePrefix = "arXiv",
    primaryClass = "astro-ph.HE",
    doi = "10.1051/0004-6361/202556986",
    journal = "Astron. Astrophys.",
    volume = "706",
    pages = "A351",
    year = "2026"
}

@article{Zech:2021emy,
    author = "Zech, Andreas and Lemoine, Martin",
    title = "{Electron-proton co-acceleration on relativistic shocks in extreme-TeV blazars}",
    eprint = "2108.12271",
    archivePrefix = "arXiv",
    primaryClass = "astro-ph.HE",
    doi = "10.1051/0004-6361/202141062",
    journal = "Astron. Astrophys.",
    volume = "654",
    pages = "A96",
    year = "2021"
}

@article{Ghisellini:2009wa,
    author = "Ghisellini, G. and Tavecchio, F.",
    title = "{Canonical high power blazars}",
    eprint = "0902.0793",
    archivePrefix = "arXiv",
    primaryClass = "astro-ph.CO",
    doi = "10.1111/j.1365-2966.2009.15007.x",
    journal = "Mon. Not. Roy. Astron. Soc.",
    volume = "397",
    pages = "985--1002",
    year = "2009"
}

@article{Kouch:2025uwz,
    author = "Kouch, Pouya M. and others",
    title = "{CAZ catalog and optical light curves of 7918 blazar-selected active galactic nuclei}",
    eprint = "2510.16584",
    archivePrefix = "arXiv",
    primaryClass = "astro-ph.HE",
    doi = "10.1051/0004-6361/202557582",
    journal = "Astron. Astrophys.",
    volume = "708",
    pages = "A382",
    year = "2026"
}

@article{Kouch:2025ipk,
    author = "Kouch, Pouya M. and Hovatta, Talvikki and Lindfors, Elina and Liodakis, Ioannis and Koljonen, Karri I. I. and Paggi, Alessandro",
    title = "{Association of the IceCube neutrinos with CAZ blazar light curves}",
    eprint = "2510.16585",
    archivePrefix = "arXiv",
    primaryClass = "astro-ph.HE",
    doi = "10.1051/0004-6361/202557584",
    journal = "Astron. Astrophys.",
    volume = "708",
    pages = "A383",
    year = "2026"
}

@article{Wang:2021unc,
    author = "Wang, Ze-Rui and Liu, Ruo-Yu and Petropoulou, Maria and Oikonomou, Foteini and Xue, Rui and Wang, Xiang-Yu",
    title = "{Unified model for orphan and multiwavelength blazar flares}",
    eprint = "2112.01739",
    archivePrefix = "arXiv",
    primaryClass = "astro-ph.HE",
    doi = "10.1103/PhysRevD.105.023005",
    journal = "Phys. Rev. D",
    volume = "105",
    number = "2",
    pages = "023005",
    year = "2022"
}

@article{Langis:2026dyj,
    author = "Langis, D. A. and Liodakis, I. and Koljonen, K. I. I. and Kouch, P. M.",
    title = "{Probing the statistical correlation of optical tidal disruption events with high-energy neutrinos}",
    eprint = "2603.20378",
    archivePrefix = "arXiv",
    primaryClass = "astro-ph.HE",
    doi = "10.1051/0004-6361/202556514",
    journal = "Astron. Astrophys.",
    volume = "709",
    pages = "A28",
    year = "2026"
}

@article{Blumenthal:1970gc,
    author = "Blumenthal, G. R. and Gould, R. J.",
    title = "{Bremsstrahlung, synchrotron radiation, and compton scattering of high-energy electrons traversing dilute gases}",
    doi = "10.1103/RevModPhys.42.237",
    journal = "Rev. Mod. Phys.",
    volume = "42",
    pages = "237--270",
    year = "1970"
}

@article{ParticleDataGroup:2020ssz,
    author = "Zyla, P. A. and others",
    collaboration = "Particle Data Group",
    title = "{Review of Particle Physics}",
    doi = "10.1093/ptep/ptaa104",
    journal = "PTEP",
    volume = "2020",
    number = "8",
    pages = "083C01",
    year = "2020"
}

@article{Luo:2026nak,
    author = "Luo, Jian-Jun and Lu, Ming-Xuan and Liang, Yun-Feng",
    title = "{On the apparent correlation between X-ray and neutrino luminosities of active galactic nuclei}",
    eprint = "2605.13588",
    archivePrefix = "arXiv",
    primaryClass = "astro-ph.HE",
    doi = "10.1016/j.jheap.2026.100639",
    journal = "JHEAp",
    volume = "53",
    pages = "100639",
    year = "2026"
}

@misc{PodlesnyiOikonomou2026Zenodo,
  doi = {10.5281/zenodo.21511152},
  url = {https://zenodo.org/doi/10.5281/zenodo.21511152},
  author = {Podlesnyi,  Egor and Oikonomou,  Foteini},
  title = {Search for neutrino emission from blazar $\gamma$-ray flares accounting for possible neutrino time delays --- Supplementary material},
  publisher = {Zenodo},
  year = {2026},
  copyright = {Creative Commons Attribution 4.0 International}
}

\appendix
    In this Appendix, we present the results of the additional tests described in Sect.~\ref{sec:caveats}.

    \begingroup 
    \setlength{\tabcolsep}{6pt} 
    \setlength\extrarowheight{2pt}
    \begin{table*}
    \centering
    \begin{tabular}{|l|l|c|c|c|c|c|c|c|c|c|}
    \hline
    Alert & Association & $s_{\nu}$ & $E_{\nu}$ [TeV] & $U_{\nu}$ [$^{{\circ}^2}$] & $t_{\nu}$ & $t^{\gamma \max}_{b}$ & $F^{\gamma}_{b}(t^{\gamma \max}_{b})$ & $\mathrm{\mathrm{TS}_{\nu}}$ & $D_{b}$ & $z_{b}$\\
    \hline
    IC170922A & PKS 0502+049 $\dagger$ & 0.631 & 264 & 7.1 & 58019 & 56918 & 0.077 & 0.6310 & 24.6 & 0.95\\
     & TXS 0506+056 &  &  &  &  & 57998 & 0.021 &  & 1.8 & 0.34\\
    \hline
    IC211208A & PKS 0735+17 $\dagger$ & 0.502 & 171 & 25.1 & 59557 & 59558 & 0.019 & 0.3124 & 4.8 & 0.42\\
    \hline
    IC220225A & PKS 0215+015 $\dagger$ & 0.378 & 154 & 27.2 & 59636 & 59648 & 0.012 & 0.2167 & 65.0 & 1.72\\
    \hline
    IC140927A & PKS 0336-01 $\dagger$ & 0.481 & 182 & 37.2 & 56927 & 56948 & 0.018 & 0.2016 & 26.3 & 0.85\\
    \hline
    IC141114A & PKS 1441+25 $\dagger$ & 0.380 & 110 & 40.6 & 56975 & 56948 & 0.008 & 0.1461 & 23.8 & 0.94\\
    \hline
    IC130125A & S5 0016+73 & 0.531 & 165 & 84.7 & 56317 & 56318 & 0.017 & 0.0979 & 19.6 & 1.78\\
    \hline
    IC140503A & 1H 1013+498 & 0.404 & 109 & 187.2 & 56781 & 56708 & 0.008 & 0.0458 & 3.1 & 0.21\\
     & S4 1030+41 $\dagger$ &  &  &  &  & 56798 & 0.005 &  & 14.6 & 1.12\\
    \hline
    IC200410A & 4C +05.64 $\dagger$ & 0.305 & 110 & 268.7 & 58950 & 57188 & 0.003 & 0.0177 & 11.6 & 1.42\\
     & TXS 1549+089 &  &  &  &  & 58958 & 0.001 &  & 1.8 & 1.01\\
     & PKS 1551+130 &  &  &  &  & 54728 & 0.013 &  & 15.6 & 1.29\\
     & PG 1553+113 &  &  &  &  & 58568 & 0.006 &  & 1.4 & 0.36\\
     & 4C +10.45 &  &  &  &  & 57068 & 0.015 &  & 16.9 & 1.23\\
    \hline
    IC150926A & 3C 279 $\dagger$ & 0.296 & 216 & 8.8 & 57292 & 57188 & 0.208 & 0.0142 & 140.2 & 0.54\\
    \hline
    IC161127A & TXS 1700+685 & 0.453 & 139 & 910.1 & 57720 & 54908 & 0.016 & 0.0078 & 8.4 & 0.30\\
     & S4 1716+68 &  &  &  &  & 55058 & 0.005 &  & 1.8 & 0.78\\
     & S4 1749+70 &  &  &  &  & 55628 & 0.022 &  & 9.1 & 0.77\\
     & S5 1803+784 &  &  &  &  & 55988 & 0.015 &  & 23.8 & 0.68\\
     & 3C 371 &  &  &  &  & 57698 & 0.008 &  & 3.8 & 0.05\\
     & S4 1842+68 &  &  &  &  & 55388 & 0.003 &  & 4.5 & 0.47\\
     & S4 1849+67 &  &  &  &  & 54818 & 0.031 &  & 48.2 & 0.66\\
    \hline
    \end{tabular}
    \caption{Same as Table~\ref{tab:top_ten_Homan} but with weights $w_{\nu}$ not including relative photon fluxes $F^{\gamma}_{b}(t^{\gamma \max}_{b})$ as described in Sect.~\ref{sec:caveats}. The local minimum of $p(\tilde{t}^{\prime}_{\mathrm{delay}}) = 0.040$ is reached at $\tilde{t}^{\prime}_{\mathrm{delay}} = 3 \times 10^{1}$~d. \label{tab:top_ten_Homan_no_flux}}
    \end{table*}
    \endgroup
    \begingroup 
    \setlength{\tabcolsep}{6pt} 
    \setlength\extrarowheight{2pt}
    \begin{table*}
    \centering
    \begin{tabular}{|l|l|c|c|c|c|c|c|c|c|c|}
    \hline
    Alert & Association & $s_{\nu}$ & $E_{\nu}$ [TeV] & $U_{\nu}$ [$^{{\circ}^2}$] & $t_{\nu}$ & $t^{\gamma \max}_{b}$ & $F^{\gamma}_{b}(t^{\gamma \max}_{b})$ & $\mathrm{\mathrm{TS}_{\nu}}$ & $D_{b}$ & $z_{b}$\\
    \hline
    IC150812B & PKS 2145+06 $\dagger$ & 0.831 & 508 & 6.5 & 57247 & 56648 & 0.003 & 0.8144 & 4.0 & 1.00\\
    \hline
    IC120605A & OL 318 $\dagger$ & 0.385 & 107 & 21.1 & 56084 & 55868 & 0.003 & 0.4544 & 19.9 & 1.41\\
     & B2 1015+35B $\dagger$ &  &  &  &  & 55148 & 0.003 &  & 3.6 & 1.23\\
    \hline
    IC111216A & TXS 0222+185 & 0.946 & 891 & 30.1 & 55911 & 55808 & 0.004 & 0.3974 & 12.3 & 2.69\\
    \hline
    IC160814A & PKS 1313-333 $\dagger$ & 0.607 & 263 & 26.3 & 57615 & 57398 & 0.033 & 0.2353 & 19.3 & 1.21\\
    \hline
    IC150926A & 3C 279 $\dagger$ & 0.296 & 216 & 8.8 & 57292 & 57188 & 0.208 & 0.1966 & 27.8 & 0.54\\
    \hline
    IC211208A & PKS 0735+17 $\dagger$ & 0.502 & 171 & 25.1 & 59557 & 59558 & 0.019 & 0.1886 & 5.8 & 0.42\\
    \hline
    IC220225A & PKS 0215+015 $\dagger$ & 0.378 & 154 & 27.2 & 59636 & 59648 & 0.012 & 0.1226 & 14.6 & 1.72\\
    \hline
    IC111208A & Mkn 421 & 0.446 & 123 & 87.1 & 55904 & 55628 & 0.014 & 0.1050 & 8.0 & 0.03\\
     & B2 1128+38 $\dagger$ &  &  &  &  & 55808 & 0.002 &  & 9.9 & 1.73\\
    \hline
    IC140927A & PKS 0336-01 $\dagger$ & 0.481 & 182 & 37.2 & 56927 & 56948 & 0.018 & 0.0859 & 26.8 & 0.85\\
    \hline
    IC170427A & RX J0011.5+0058 $\dagger$ & 0.383 & 155 & 40.3 & 57870 & 57578 & 0.012 & 0.0844 & 17.3 & 1.49\\
     & S3 0013-00 &  &  &  &  & 56018 & 0.008 &  & 15.0 & 1.58\\
    \hline
    \end{tabular}
    \caption{Same as Table~\ref{tab:top_ten_Rodrigues} but with weights $w_{\nu}$ not including relative photon fluxes $F^{\gamma}_{b}(t^{\gamma \max}_{b})$ as described in Sect.~\ref{sec:caveats}. The local minimum of $p(\hat{t}^{\prime}_{\mathrm{delay}}) = 0.079$ is reached at $\hat{t}^{\prime}_{\mathrm{delay}} = 9.8 \times 10^{2}$~d. \label{tab:top_ten_Rodrigues_no_flux}}
    \end{table*}
    \endgroup

    \begin{figure}[t]
        \centering
        \includegraphics[width=0.45\textwidth]{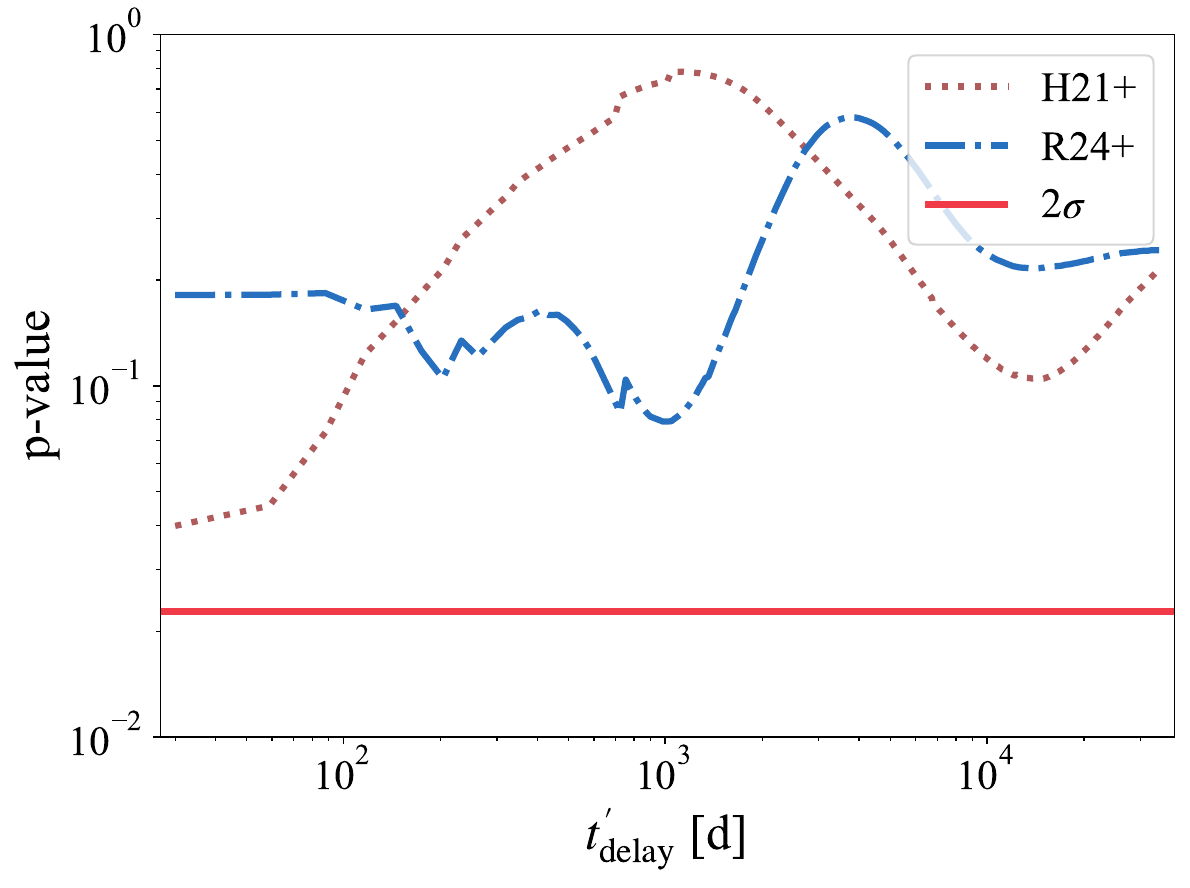}
        \caption{Same as Fig.~\ref{fig:p_value_plot} but for the tests not including relative photon fluxes $F^{\gamma}_{b}(t^{\gamma \max}_{b})$ in the weights $w_{\nu}$ as described in Sect.~\ref{sec:caveats}; $p_{\mathrm{post-trial\, H21+}} = 17$\% ($0.94\sigma$) and $p_{\mathrm{post-trial\, R24+}} = 29$\% ($0.55\sigma$). \label{fig:p_value_plot_no_flux}}
    \end{figure}
    
    \begingroup 
    \setlength{\tabcolsep}{6pt} 
    \setlength\extrarowheight{2pt}
    \begin{table*}
    \centering
    \begin{tabular}{|l|l|c|c|c|c|c|c|c|c|c|}
    \hline
    Alert & Association & $s_{\nu}$ & $E_{\nu}$ [TeV] & $U_{\nu}$ [$^{{\circ}^2}$] & $t_{\nu}$ & $t^{\gamma \max}_{b}$ & $F^{\gamma}_{b}(t^{\gamma \max}_{b})$ & $\mathrm{\mathrm{TS}_{\nu}}$ & $D_{b}$ & $z_{b}$\\
    \hline
    IC150812B & PKS 2145+06 $\dagger$ & 0.831 & 508 & 6.5 & 57247 & 56648 & 0.003 & 0.8310 & 31.0 & 0.99\\
    \hline
    IC221223A & B2 2308+34 $\dagger$ & 0.795 & 353 & 7.5 & 59936 & 59588 & 0.028 & 0.7950 & 22.8 & 1.82\\
    \hline
    IC190730A & PKS 1502+106 & 0.670 & 298 & 14.2 & 58695 & 57218 & 0.079 & 0.6700 & 65.2 & 1.84\\
    \hline
    IC170922A & PKS 0502+049 $\dagger$ & 0.631 & 264 & 7.1 & 58019 & 56918 & 0.077 & 0.6310 & 24.6 & 0.95\\
     & TXS 0506+056 &  &  &  &  & 57998 & 0.021 &  & 1.8 & 0.34\\
    \hline
    IC150831A & NRAO 140 $\dagger$ & 0.579 & 181 & 10.4 & 57265 & 55568 & 0.008 & 0.5790 & 30.5 & 1.26\\
    \hline
    IC200916A & 4C +14.23 $\dagger$ & 0.322 & 110 & 10.2 & 59109 & 55148 & 0.055 & 0.3220 & 13.9 & 1.04\\
    \hline
    IC121115A & PKS 1502+106 $\dagger$ & 0.319 & 116 & 10.2 & 56246 & 54878 & 0.078 & 0.3190 & 65.2 & 1.84\\
    \hline
    IC150926A & 3C 279 $\dagger$ & 0.296 & 216 & 8.8 & 57292 & 57188 & 0.208 & 0.2960 & 140.2 & 0.54\\
    \hline
    IC230708A & OT 081 & 0.560 & 208 & 56.6 & 60133 & 57578 & 0.074 & 0.1545 & 102.4 & 0.32\\
     & TXS 1811+062 $\dagger$ &  &  &  &  & 59558 & 0.001 &  & 4.4 & 0.54\\
    \hline
    IC211123A & OT 081 $\dagger$ & 0.356 & 142 & 42.1 & 59542 & 57578 & 0.074 & 0.1319 & 102.4 & 0.32\\
    \hline
    \end{tabular}
    \caption{Same as Table~\ref{tab:top_ten_Homan} but with weights $\omega_{\nu}$ from Eq.~(\ref{eq:alternative_weights}) described in Sect.~\ref{sec:caveats}. Mind that $t^{\gamma \max}_{b}$ and $F^{\gamma}_{b}(t^{\gamma \max}_{b})$ do not enter into $\omega_{\nu}$ but are kept to have the table format unchanged. The local minimum of $p(\tilde{t}^{\prime}_{\mathrm{delay}}) = 0.011$ is reached at $\tilde{t}^{\prime}_{\mathrm{delay}} = 2.7 \times 10^{4}$~d. \label{tab:top_ten_Homan_68}}
    \end{table*}
    \endgroup
    \begingroup 
    \setlength{\tabcolsep}{6pt} 
    \setlength\extrarowheight{2pt}
    \begin{table*}
    \centering
    \begin{tabular}{|l|l|c|c|c|c|c|c|c|c|c|}
    \hline
    Alert & Association & $s_{\nu}$ & $E_{\nu}$ [TeV] & $U_{\nu}$ [$^{{\circ}^2}$] & $t_{\nu}$ & $t^{\gamma \max}_{b}$ & $F^{\gamma}_{b}(t^{\gamma \max}_{b})$ & $\mathrm{\mathrm{TS}_{\nu}}$ & $D_{b}$ & $z_{b}$\\
    \hline
    IC190730A & PKS 1502+106 & 0.670 & 298 & 14.2 & 58695 & 57218 & 0.079 & 0.6700 & 22.1 & 1.83\\
    \hline
    IC170922A & PKS 0502+049 $\dagger$ & 0.631 & 264 & 7.1 & 58019 & 56918 & 0.077 & 0.6310 & 29.8 & 0.95\\
    \hline
    IC120515A & OP 313 & 0.613 & 194 & 13.3 & 56063 & 54698 & 0.012 & 0.6130 & 31.5 & 1.00\\
    \hline
    IC130127A & PKS 2335-027 $\dagger$ & 0.610 & 235 & 10.5 & 56319 & 55268 & 0.007 & 0.6100 & 16.4 & 1.07\\
    \hline
    IC160806A & PKS B0802-010 $\dagger$ & 0.578 & 219 & 7.7 & 57607 & 55808 & 0.007 & 0.5780 & 19.3 & 1.39\\
    \hline
    IC200916A & 4C +14.23 $\dagger$ & 0.322 & 110 & 10.2 & 59109 & 55148 & 0.055 & 0.3220 & 9.7 & 1.04\\
    \hline
    IC121115A & PKS 1502+106 $\dagger$ & 0.319 & 116 & 10.2 & 56246 & 54878 & 0.078 & 0.3190 & 22.1 & 1.83\\
    \hline
    IC211208A & PKS 0735+17 $\dagger$ & 0.502 & 171 & 25.1 & 59557 & 59558 & 0.019 & 0.3124 & 5.8 & 0.42\\
    \hline
    IC160727A & 4C +14.23 $\dagger$ & 0.296 & 105 & 14.9 & 57596 & 55148 & 0.055 & 0.2960 & 9.7 & 1.04\\
    \hline
    IC150926A & 3C 279 $\dagger$ & 0.296 & 216 & 8.8 & 57292 & 57188 & 0.208 & 0.2960 & 27.8 & 0.54\\
    \hline
    \end{tabular}
    \caption{Same as Table~\ref{tab:top_ten_Rodrigues} but with weights $\omega_{\nu}$ from Eq.~(\ref{eq:alternative_weights}) described in Sect.~\ref{sec:caveats}. Mind that $t^{\gamma \max}_{b}$ and $F^{\gamma}_{b}(t^{\gamma \max}_{b})$ do not enter into $\omega_{\nu}$ but are kept to have the table format unchanged. The local minimum of $p(\hat{t}^{\prime}_{\mathrm{delay}}) = 0.022$ is reached at $\hat{t}^{\prime}_{\mathrm{delay}} = 1.1 \times 10^{4}$~d. \label{tab:top_ten_Rodrigues_68}}
    \end{table*}
    \endgroup

    \begin{figure}[t]
        \centering
        \includegraphics[width=0.45\textwidth]{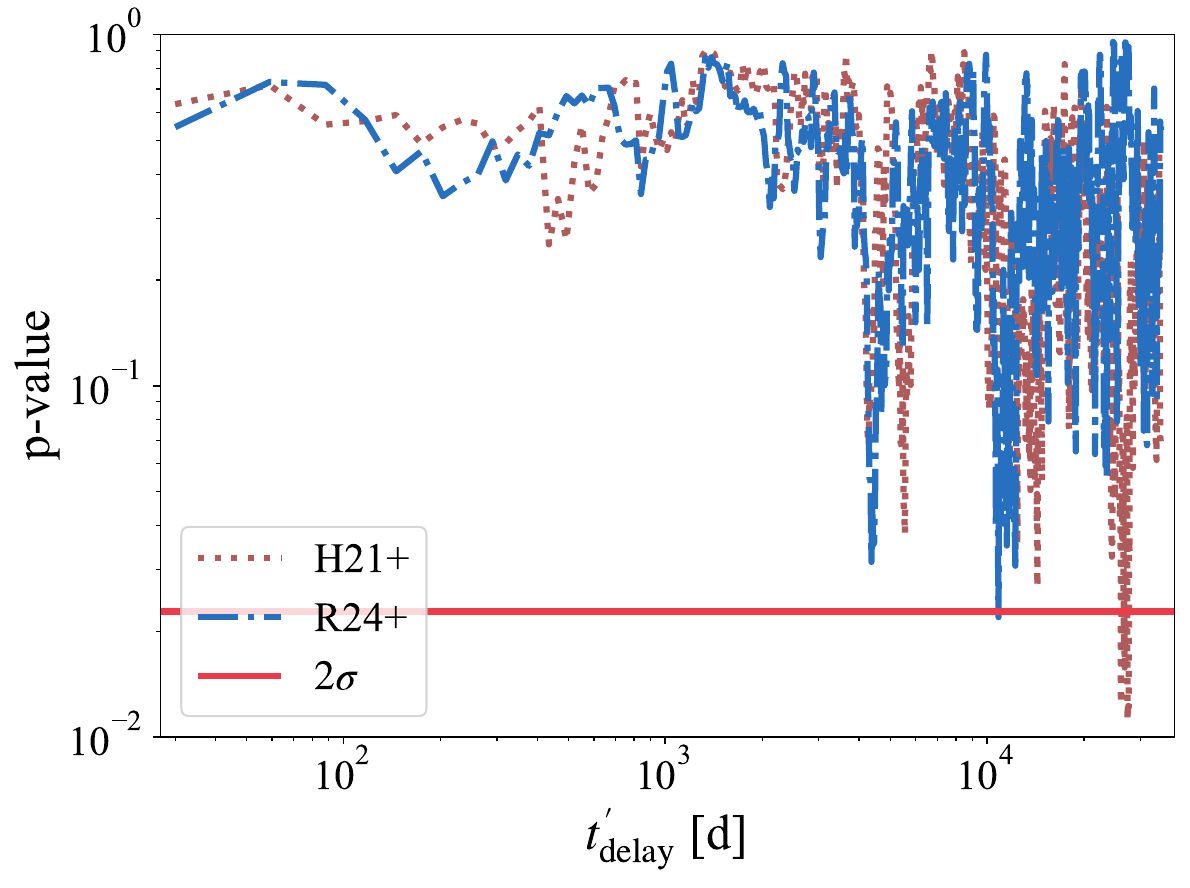}
        \caption{Same as Fig.~\ref{fig:p_value_plot} but but with weights $\omega_{\nu}$ from Eq.~(\ref{eq:alternative_weights}) described in Sect.~\ref{sec:caveats}; $p_{\mathrm{post-trial\, H21+}} = 27$\% ($0.62\sigma$) and $p_{\mathrm{post-trial\, R24+}} = 47$\% ($0.09\sigma$). \label{fig:p_value_plot_68}}
    \end{figure}
    
    \begingroup 
    \setlength{\tabcolsep}{6pt} 
    \setlength\extrarowheight{2pt}
    \begin{table*}
    \centering
    \begin{tabular}{|l|l|c|c|c|c|c|c|c|c|c|c|}
    \hline
    Alert & Association & $s_{\nu}$ & $E_{\nu}$ [TeV] & $U_{\nu}$ [$^{{\circ}^2}$] & $t_{\nu}$ & $t^{\gamma \max}_{b}$ & $F^{\gamma}_{b}(t^{\gamma \max}_{b})$ & $F^{X}_{b}$ & $\mathrm{\mathrm{TS}_{\nu}}$ & $D_{b}$ & $z_{b}$\\
    \hline
    IC120523B & 3C 454.3 & 0.490 & 168 & 47.2 & 56071 & 55538 & 1.000 & 0.691 & 0.0779 & 45.3 & 0.86\\
    \hline
    IC150926A & 3C 279 $\dagger$ & 0.296 & 216 & 8.8 & 57292 & 57188 & 0.208 & 0.560 & 0.0345 & 140.2 & 0.54\\
    \hline
    IC180608A & PKS 0440-00 $\dagger$ & 0.396 & 158 & 11.3 & 58278 & 56498 & 0.047 & 0.065 & 0.0011 & 3.5 & 0.45\\
    \hline
    IC170922A & PKS 0502+049 $\dagger$ & 0.631 & 264 & 7.1 & 58019 & 56918 & 0.077 & 0.029 & 0.0010 & 24.6 & 0.95\\
     & TXS 0506+056 &  &  &  &  & 57998 & 0.021 & 0.017 & & 1.8 & 0.34\\
    \hline
    IC150812B & PKS 2145+06 $\dagger$ & 0.831 & 508 & 6.5 & 57247 & 56648 & 0.003 & 0.297 & 0.0007 & 31.0 & 0.99\\
    \hline
    IC221223A & B2 2308+34 $\dagger$ & 0.795 & 353 & 7.5 & 59936 & 59588 & 0.028 & 0.030 & 0.0005 & 22.8 & 1.82\\
    \hline
    IC150625A & PKS 0440-00 & 0.286 & 112 & 48.2 & 57199 & 56498 & 0.047 & 0.065 & 0.0002 & 3.5 & 0.45\\
    \hline
    IC120515A & OP 313 & 0.613 & 194 & 13.3 & 56063 & 54698 & 0.012 & 0.070 & 0.0002 & 24.2 & 1.00\\
    \hline
    IC140927A & PKS 0336-01 $\dagger$ & 0.481 & 182 & 37.2 & 56927 & 56948 & 0.018 & 0.085 & 0.0002 & 26.3 & 0.85\\
    \hline
    IC160727A & 4C +14.23 $\dagger$ & 0.296 & 105 & 14.9 & 57596 & 55148 & 0.055 & 0.029 & 0.0002 & 13.9 & 1.04\\
    \hline
    \end{tabular}
    \caption{Same as Table~\ref{tab:top_ten_Homan} but with an extra multiplying factor $F_{b}^{X}$ of the radio flux of the associated blazar in the statistical weight $w_{\nu}$ as described in Sect.~\ref{sec:caveats}. The relative radio flux $F_{b}^{X}$ in the X radio band (between 8 GHz and 12 GHz) of the associated source is shown in the added column.  The local minimum of $p(\tilde{t}^{\prime}_{\mathrm{delay}}) = 0.026$ is reached at $\tilde{t}^{\prime}_{\mathrm{delay}} = 7 \times 10^{3}$~d. \label{tab:top_ten_Homan_radio}}
    \end{table*}
    \endgroup

    \begin{figure}[t]
        \centering
        \includegraphics[width=0.45\textwidth]{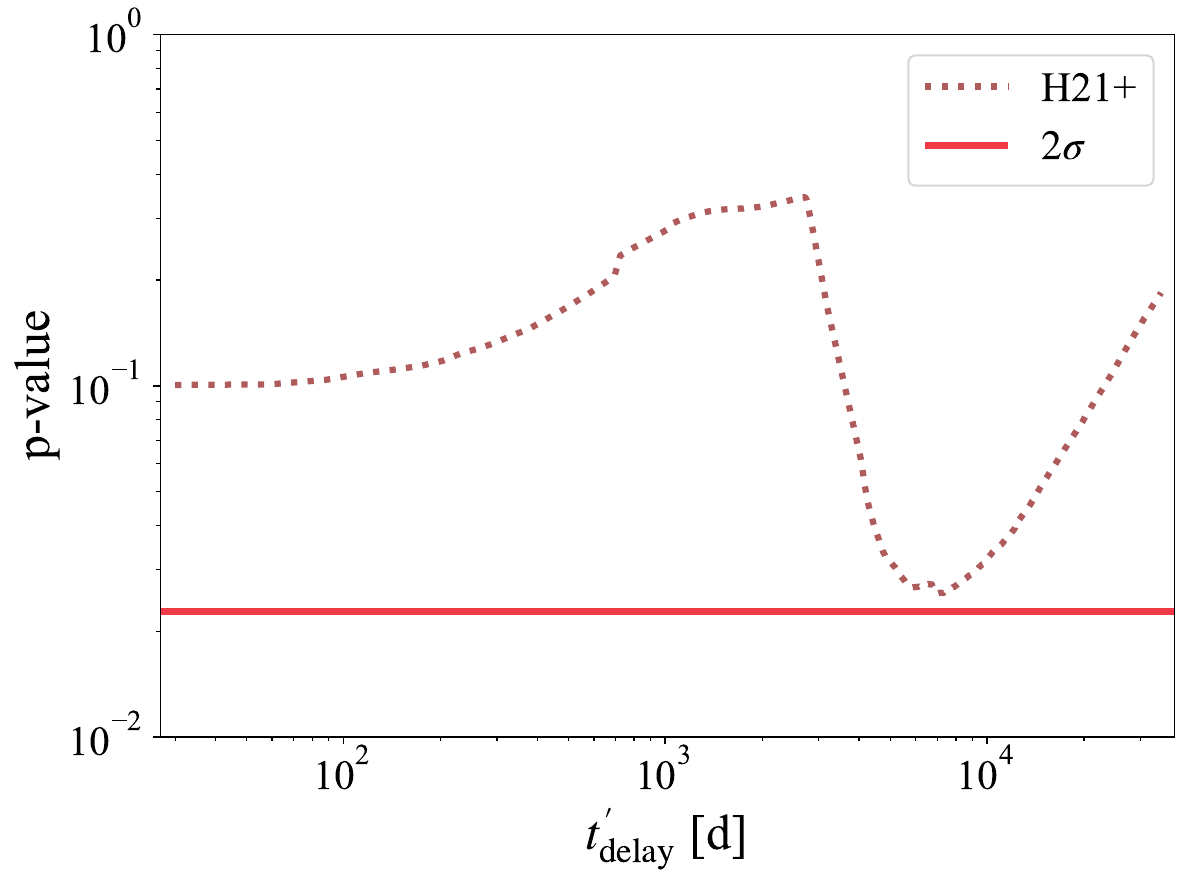}
        \caption{Same as Fig.~\ref{fig:p_value_plot} but for the test with the H21+ source list with extra radio weights $F_{b}^{X}$ as described in Sect.~\ref{sec:caveats}; $p_{\mathrm{post-trial\, H21+}} = 11$\% ($1.25\sigma$). \label{fig:p_value_plot_homan_radio}}
    \end{figure}

\end{document}